\documentclass[trackchanges]{aastex701}
\usepackage{rotating}
\usepackage{adjustbox}
\usepackage{graphicx}
\usepackage{capt-of}  
\newcommand{\Nv}{\ion{N}{5}}
\newcommand{\Niv}{\ion{N}{4}]}
\newcommand{\Niii}{\ion{N}{3}]}
\newcommand{\Civ}{\ion{C}{4}}
\newcommand{\Ciii}{\ion{C}{3}]}
\newcommand{\Mgii}{\ion{Mg}{2}}
\newcommand{\Feii}{\ion{Fe}{2}}
\newcommand{\Heii}{\ion{He}{2}}
\newcommand{\Aliii}{\ion{Al}{3}}

\newcommand{\Oiii}{\ion{O}{3}}
\defcitealias{Zhai2026PaperI}{Paper I}

\def\IhepCAS{Key Laboratory for Particle Astrophysics, Institute of High Energy Physics, Chinese Academy of Sciences, 19B Yuquan Road, Beijing 100049, P. R. China}

\def\UCASast{School of Astronomy and Space Science, University of Chinese Academy of Sciences, 19A Yuquan Road, Beijing 100049, P. R. China}

\def\naocOptical{National Astronomical Observatories, Chinese Academy of Sciences (CAS), Beijing 100101, P. R. China}
\def\SpaceTechnology{Technology and Engineering Center for Space Utilization, Chinese Academy of Sciences, Beijing 100094, P. R. China}

\begin{document}

\title{Nitrogen-Loud Quasars from the Dark Energy Spectroscopic Instrument. II. Broad-Line Region Metallicity and Relative Nitrogen Enrichment}

\author[0009-0005-4152-2088]{Shuo Zhai}
\email{}
\affiliation{\naocOptical}

\author[0000-0001-9457-0589]{Wei-Jian Guo}
\email{}
\affiliation{\naocOptical}

\author[0000-0003-4280-7673]{Yong-Jie Chen}
\email{}
\affiliation{\SpaceTechnology}

\author[0000-0002-0389-9264]{Haining Li}
\email{}
\affiliation{\naocOptical}
\affiliation{\UCASast}––

\author[0000-0001-9449-9268]{Jian-Min Wang}
\email{}
\affiliation{\IhepCAS}
\affiliation{\naocOptical}
\affiliation{\UCASast}


\author[0000-0002-8980-945X]{Gang Zhao}
\email{}
\affiliation{\naocOptical}
\affiliation{\UCASast}

\begin{abstract}

Whether the unusually strong nitrogen emission in nitrogen-loud (N-loud)
quasars reflects high overall metallicity, enhanced nitrogen abundance relative
to other elements, or both has long been debated. We analyze the broad-line
region (BLR) abundances of 121 N-loud quasars at
$2.13 \leq z \leq 3.90$ selected from the Dark Energy Spectroscopic Instrument
Data Release 1 and construct a control sample of normal quasars matched in
redshift, continuum luminosity, and virial black hole mass. Within the N-loud
sample, metallicities inferred from \Nv/\Civ\ span $\sim3$-$50\,Z_\odot$ and are
systematically higher than those inferred from the nitrogen-independent
(\ion{Si}{4}+\ion{O}{4}])/\Civ\ and \Aliii/\Civ\ diagnostics, which agree
closely and mainly span $\sim1$-$20\,Z_\odot$. Compared with the matched
controls, the N-loud quasars show systematically higher metallicities in both
\Nv/\Civ\ and (\ion{Si}{4}+\ion{O}{4}])/\Civ, with median values approximately
three times those of the controls. Notably, the discrepancy between the
metallicities inferred from \Nv/\Civ\ and (\ion{Si}{4}+\ion{O}{4}])/\Civ\ becomes
more pronounced toward the high-metallicity end of the N-loud sample,
suggesting additional nitrogen enrichment beyond the overall metal enrichment.
Together, these results indicate that N-loud
quasars have both high overall BLR metallicity and enhanced relative nitrogen
abundance, suggesting that relative nitrogen abundance is at least partially
decoupled from overall metallicity. N-loud quasars therefore provide a
high-metallicity laboratory for understanding how nuclear environments can
produce unusual abundance patterns and offer a complementary view of how
nitrogen enrichment arises across cosmic time.

\end{abstract}

\keywords{\uat{Active galactic nuclei}{16} --- \uat{Active galaxies}{17} ---
\uat{Quasars}{1319} --- \uat{Chemical abundances}{224} --- \uat{Metallicity}{1031}}

\section{Introduction}
\label{sec:intro}

Chemical abundances in galaxies and active galactic nuclei (AGNs) encode the
cumulative effects of star formation and gas cycling, providing important
constraints on galaxy evolution and chemical enrichment
\citep{Tinsley1980,Pagel1997,HamannFerland1999,MaiolinoMannucci2019}. The
prominent broad emission lines in quasars provide access to the chemical
abundances of gas in the broad-line region (BLR) and have been widely used to
study quasar metallicities over a broad range of redshifts
\citep{HamannFerland1993,HamannFerland1999,Hamann2002,Dietrich2003,Nagao2006}.
Nitrogen lines, particularly \Nv~$\lambda1240$, are especially useful in these
studies. In metal-rich \ion{H}{2} regions, N/O is observed to increase with O/H,
a behavior commonly attributed to secondary nitrogen production from
pre-existing C and O through the carbon--nitrogen--oxygen (CNO) cycle
\citep{Shields1976,Tinsley1980,VilaCostasEdmunds1993,vanZee1998}. When secondary
production dominates, this behavior is approximately described by
$\mathrm{N/O} \propto \mathrm{O/H}$, or equivalently $\mathrm{N/H} \propto (\mathrm{O/H})^2$. This
secondary-nitrogen scaling provides the basis for interpreting nitrogen-based
BLR diagnostics in terms of overall metallicity
\citep{HamannFerland1993,Hamann2002}.

Within this framework, the exceptionally strong nitrogen emission of nitrogen-loud (N-loud)
quasars can imply BLR metallicities well above the
$\sim4$-$5\,Z_\odot$ values typically inferred for luminous quasars
\citep{Dietrich2003,Nagao2006,Matsuoka2011}. Such unusually high estimates
raise the possibility that strong nitrogen emission may reflect not only
overall metal enrichment but also an enhanced nitrogen abundance relative to
other elements. Recent James Webb Space Telescope (JWST) observations have identified several high-redshift
systems with unusually high N/O at relatively low oxygen abundance
\citep{Cameron2023,Isobe2023,MarquesChaves2024,Topping2024}. These observations
demonstrate that relative nitrogen abundance can depart from the abundance
pattern expected from overall metallicity alone.

The abundance interpretation of N-loud quasars has long centered on whether
their strong nitrogen emission reflects overall metal enrichment or a selective
enhancement of nitrogen relative to other elements. Detailed photoionization
modeling of Q0353$-$383 inferred a BLR metallicity of $\sim15\,Z_\odot$ under
the standard secondary-nitrogen scaling \citep{Baldwin2003}. \citet{Batra2014}
further analyzed 43 of the most extreme N-loud quasars using
multiple BLR abundance diagnostics spanning different ionization states and
inferred a mean metallicity of $\sim5.5\,Z_\odot$, with values reaching
$\sim18\,Z_\odot$, arguing that these quasars are overall metal rich rather than
characterized by nitrogen enhancement alone. In contrast, \citet{Jiang2008}
found that many non-nitrogen emission properties of N-loud quasars are similar
to those of normal quasars and suggested that their unusually strong nitrogen
emission may primarily reflect enhanced relative nitrogen abundance.
\citet{Araki2012} and \citet{Matsuoka2017} used near-infrared spectroscopy to measure
rest-frame optical narrow emission lines and used the strength of
[\ion{O}{3}]~$\lambda5007$ to constrain extreme metal enrichment in the
narrow-line region (NLR). At very high metallicity, enhanced metal cooling
lowers the gas temperature and can suppress [\ion{O}{3}] emission, whereas the
observed [\ion{O}{3}] strengths in these N-loud quasars are comparable to those
of normal quasars, providing no evidence for comparably extreme enrichment in
their NLRs. Thus, it remains unclear whether the unusually strong nitrogen
emission in N-loud quasars primarily reflects overall metal enrichment,
additional nitrogen enhancement, or a combination of both.

Previous studies have not directly separated these two effects within the BLR.
Studies supporting high BLR metallicities relied primarily on emission-line
ratios involving \Nv~$\lambda1240$, \Niv~$\lambda1486$, and
\Niii~$\lambda1750$, so their inferred metallicities remain tied to the adopted
secondary-nitrogen scaling \citep{Baldwin2003,Batra2014}. The rest-frame optical
studies discussed above provide complementary constraints from the NLR, but the
BLR and NLR probe gas on very different spatial scales and under different
physical conditions. Although correlations between BLR and NLR metallicity
indicators have been reported \citep{Du2014}, abundance constraints from the two
regions are therefore not directly interchangeable. In addition, even within
the same BLR spectrum, different abundance diagnostics respond differently to
gas density, ionization state, the ionizing spectral energy distribution (SED), and the distribution of BLR
clouds, potentially producing diagnostic-dependent metallicity offsets
\citep{Hamann2002,Nagao2006,Batra2014}. High-ionization ratios such as
\Nv/\Civ\ can also depend on BLR density, ionizing flux, and kinematic
structure, while \Civ\ blueshift can systematically affect ratios normalized by
\Civ~\citep{Temple2021,Lai2022}. A direct comparison of nitrogen-based and
nitrogen-independent BLR diagnostics within a controlled quasar sample is
therefore needed to distinguish overall metal enrichment from additional
nitrogen enhancement.

To distinguish these two effects more directly, we apply both nitrogen-based
and nitrogen-independent BLR abundance diagnostics consistently to N-loud
quasars and matched normal quasars and compare the resulting measurements. We
select objects with high-quality rest-frame UV spectra from the N-loud quasar
catalog presented in \citet[hereafter Paper I]{Zhai2026PaperI}, which was
constructed using data from the Dark Energy Spectroscopic Instrument (DESI)
Data Release 1 (DR1; \citealt{DESI2025}). For each N-loud quasar, we construct a
control sample of normal quasars matched in redshift, continuum luminosity, and
black hole mass. The primary matching uses \Civ-based black hole masses, with an
additional \Mgii-based control sample used to test the robustness of the results
to the choice of black hole mass estimator. The primary abundance comparison
focuses on \Nv/\Civ\ as the nitrogen-based diagnostic and
(\ion{Si}{4}+\ion{O}{4}])/\Civ\ as the nitrogen-independent diagnostic,
allowing us to assess overall metal enrichment and test whether the nitrogen
emission requires an additional enhancement in nitrogen abundance.

The paper is organized as follows. In Section~\ref{sec:samples}, we describe the
N-loud quasar sample and the construction of the \Civ- and \Mgii-based
control samples. In Section~\ref{sec:fitting}, we present the emission-line
measurements and the BLR abundance diagnostics adopted in this work. In
Section~\ref{sec:results}, we first examine the consistency among different
metallicity diagnostics within the N-loud sample, then compare the N-loud
quasars with the \Civ- and \Mgii-based matched controls, and finally investigate
the dependence of the line-ratio excesses on black hole mass and Eddington
ratio. In Section~\ref{sec:discussion}, we discuss the implications for overall
metallicity and relative nitrogen enrichment, possible enrichment scenarios
and their connection to black hole accretion, and comparisons with
nitrogen-enriched systems at high redshift. Our main conclusions are summarized in
Section~\ref{sec:sum}. Throughout this paper, we adopt a flat $\Lambda$CDM
cosmology with $H_0=70~\mathrm{km~s^{-1}~Mpc^{-1}}$, $\Omega_m=0.30$, and
$\Omega_\Lambda=0.70$.
\section{Sample Construction and Properties}
\label{sec:samples}

To characterize the BLR abundance properties of N-loud quasars and compare
them with those of normal quasars, we construct the analysis and control
samples used throughout this work. We first select an N-loud subsample with
complete rest-frame UV coverage and sufficient spectral quality for uniform
emission line measurements (Section~\ref{sec:nloud_sample}). We then construct
a primary control sample matched in redshift, continuum luminosity, and C
IV-based virial black hole mass (Section~\ref{sec:civ_control_sample}), together
with an independent Mg II-based control sample to assess the sensitivity of
the results to the adopted black hole mass estimator
(Section~\ref{sec:mgii_control_sample}). Finally, in
Section~\ref{sec:sample_properties}, we examine the matching quality and
summarize the global properties of the resulting samples.

\subsection{N-loud Sample}
\label{sec:nloud_sample}

We select the analysis sample from the 2,198 N-loud quasars identified in
\citetalias{Zhai2026PaperI}, applying additional criteria to ensure reliable BLR abundance
measurements. We require full rest-frame spectral coverage from 1150 to
2000~$\mathrm{\mathring{A}}$, encompassing the emission lines used in the
abundance diagnostics adopted in this work, including \Nv,
\ion{Si}{4}+\ion{O}{4}], \Civ, \Oiii], \Niii, \Aliii, and \Ciii. For DESI
spectra, this wavelength coverage corresponds to a redshift range of
$2.13\leq z\leq3.90$.

We further impose spectral quality criteria. \Niii\ is detected much more
frequently than \Niv\ among N-loud quasars (Jiang et al.\ 2008; \citetalias{Zhai2026PaperI}), and
we therefore adopt \Niii\ for the emission line quality selection. \Niii\ is
also a relatively weak intercombination line and can be affected by blending
and noise. To ensure reliable measurements, we require \Niii\ to have
$\mathrm{S/N}\geq10$, with the \Niii\ $\mathrm{S/N}$ defined as the
continuum-subtracted flux integrated over the rest-frame
1745-1755~$\mathrm{\mathring{A}}$ interval divided by the propagated flux
uncertainty over the same interval. We also require a rest-frame equivalent
width (EW) $>5$~$\mathrm{\mathring{A}}$, using the EW measurements reported in
\citetalias{Zhai2026PaperI}. We further require $\mathrm{S/N}>5$ in both the rest-frame
1430-1460~$\mathrm{\mathring{A}}$ and 1675-1725~$\mathrm{\mathring{A}}$
continuum windows, where the continuum $\mathrm{S/N}$ is defined as the mean
flux density divided by the mean spectral uncertainty within each window.
Spectra showing broad or narrow absorption that significantly affects the
emission lines used in the abundance analysis are excluded through visual
inspection. These criteria yield a final sample of 121 N-loud quasars.

\subsection{C IV-based Control Sample}
\label{sec:civ_control_sample}

To compare the abundance properties of N-loud quasars with those of normal
quasars, we construct a control sample matched in redshift, continuum
luminosity, and black hole mass. Matching in redshift reduces redshift-dependent
observational and sample-selection effects. Continuum luminosity and black
hole mass are also included since metallicity-sensitive emission line ratios
have been reported to correlate with both quantities
\citep{Warner2003,Nagao2006,Matsuoka2011,Xu2018}.

Candidate normal quasars are drawn from the DESI DR1 quasar catalog after
excluding the N-loud quasars identified in \citetalias{Zhai2026PaperI}. We restrict the candidates
to the same redshift range as the N-loud sample, $2.13\leq z\leq3.90$, and
apply the same continuum $\mathrm{S/N}$ requirement of $>5$ in both the
rest-frame 1430-1460~$\mathrm{\mathring{A}}$ and
1675-1725~$\mathrm{\mathring{A}}$ windows. We adopt \Civ-based virial black
hole masses for the primary matching because \Civ\ is covered across the full
redshift range of the N-loud sample. As cataloged virial black hole mass
estimates are not available for DESI DR1 quasars over this redshift range, we
estimate \Civ-based black hole masses consistently for the N-loud and
candidate control quasars following the procedure described in
Appendix~\ref{app:civ_bh_mass}.

For each N-loud quasar, we search for controls in the three-dimensional space
of $(z,\log L_{1350},\log M_{\rm BH,C\,IV})$. Candidate pairs are required to
satisfy $|\Delta z|<0.10$, $|\Delta\log L_{1350}|<0.20$ dex, and
$|\Delta\log M_{\rm BH,C\,IV}|<0.20$ dex. These tolerances were chosen
empirically to allow most N-loud quasars to be matched with three to five
control quasars while maintaining close agreement in all three matching
parameters. Within these limits, candidates are ranked by their Euclidean
distance after each matching variable is scaled by its standard deviation in
the combined N-loud and eligible control samples. Controls are assigned
without replacement in successive rounds, with at most one control assigned
to each N-loud quasar per round. In each round, N-loud quasars with fewer
remaining eligible controls are processed first, and the nearest available
control is assigned. This procedure is repeated until each N-loud quasar has
up to five controls or no eligible controls remain. After matching, we
visually inspect the matched control spectra and exclude quasars showing broad
or narrow absorption that significantly affects the emission lines used in
the abundance analysis.

Following this inspection, the final \Civ-based control sample contains 420
unique control quasars matched to 120 of the 121 N-loud quasars, with a median
of three controls per matched N-loud quasar. The single unmatched N-loud
quasar is excluded from the matched-sample comparisons.

\subsection{Mg II-based Control Sample}
\label{sec:mgii_control_sample}

Given that \Civ-based virial black hole masses can be affected by non-virial
contributions associated with winds and outflows, potentially introducing
systematic biases in single-epoch mass estimates
\citep{Shen2008,Denney2012,Coatman2017}, we construct an independent control
sample based on \Mgii\ virial masses. Candidate normal quasars are selected
following the same criteria as for the \Civ-based control sample, but are
restricted to $2.13\leq z\leq2.27$ to ensure \Mgii\ coverage. \Mgii-based
virial black hole masses are estimated consistently for the N-loud and
candidate control quasars following the procedure described in
Appendix~\ref{app:mgii_bh_mass}.

For each N-loud quasar, controls are matched in
$(z,\ \log L_{3000},\ \log M_{\rm BH,Mg\,II})$ following the same procedure
as for the \Civ-based control sample, with $|\Delta z|<0.10$,
$|\Delta\log L_{3000}|<0.20$ dex, and
$|\Delta\log M_{\rm BH,Mg\,II}|<0.20$ dex. Up to five controls are assigned
to each N-loud quasar without replacement.

The final \Mgii-based control sample contains 68 unique control quasars
matched to the 16 N-loud quasars with reliable \Mgii\ measurements. The
median number of control quasars per N-loud quasar is four.
In addition, we compare the \Civ- and \Mgii-based black hole
masses for the 16 N-loud quasars and their control quasars drawn from both
matched-control samples, including only objects with both measurements
available (Appendix~\ref{app:civ_mgii_bh_mass_comparison};
Figure~\ref{fig:civ_mgii_bh_mass_comparison}). The comparison shows small
median offsets in both samples, although appreciable object-to-object scatter
remains and may affect control selection. We therefore use the \Mgii-based
control sample as an alternative matching scheme to test the robustness of
our abundance results against the choice of black hole mass estimator.

\subsection{Sample Properties}
\label{sec:sample_properties}

\begin{figure*}[t]
\centering
\includegraphics[width=1.0\textwidth]{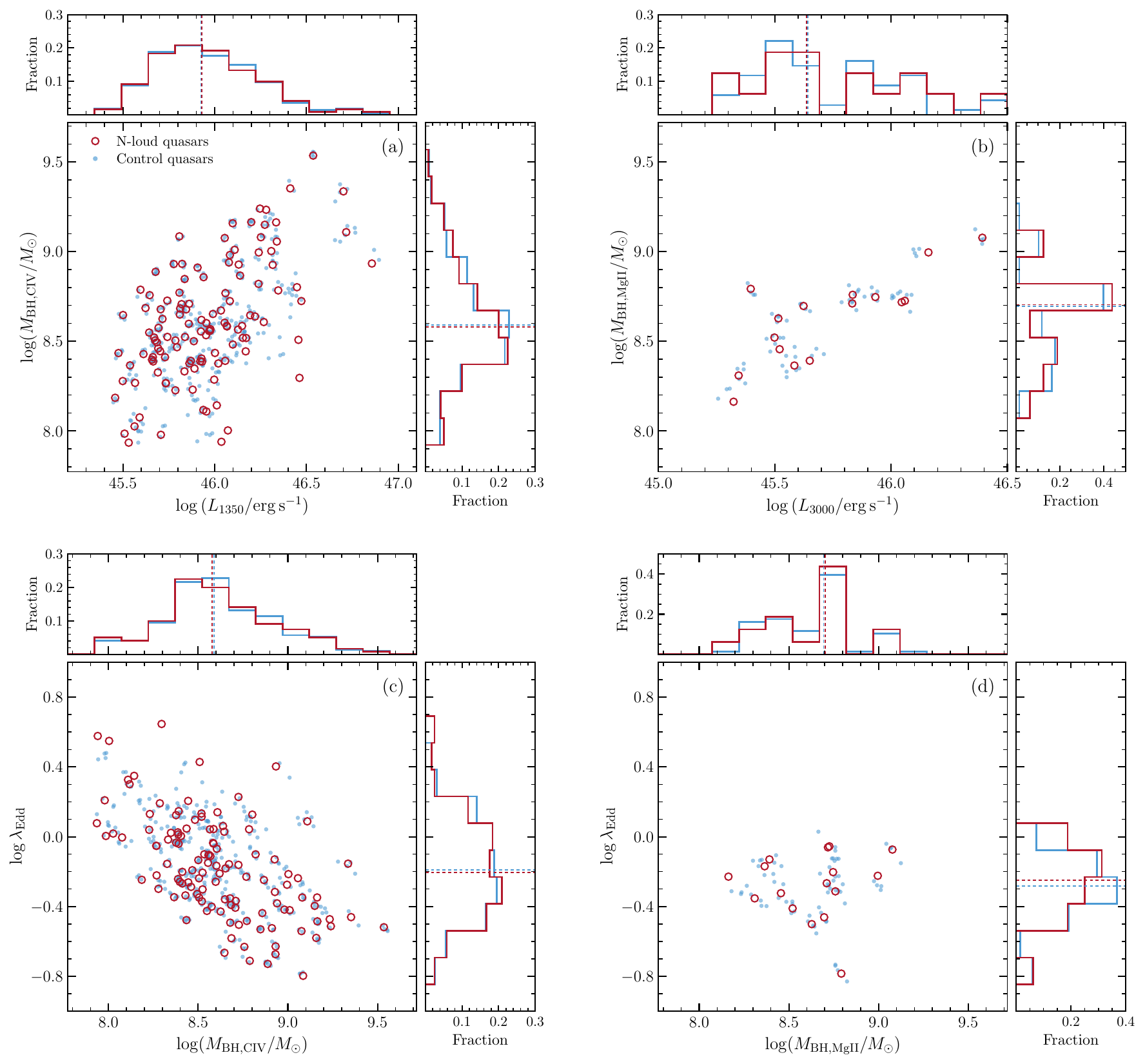}
\caption{Matching quality and global properties of the N-loud quasars and
their matched normal-quasar controls. Red open circles represent the N-loud
quasars, and blue filled circles represent the matched controls. Panels (a)
and (b) show the distributions in the $L_{1350}$--$M_{\rm BH,C\,IV}$ and
$L_{3000}$--$M_{\rm BH,Mg\,II}$ planes for the \Civ- and \Mgii-based samples,
respectively. Panels (c) and (d) show the corresponding distributions in the
$M_{\rm BH,C\,IV}$--$\lambda_{\rm Edd}$ and
$M_{\rm BH,Mg\,II}$--$\lambda_{\rm Edd}$ planes. The upper and right marginal
histograms show the normalized distributions of the corresponding quantities,
with the red and blue dashed lines marking the medians of the N-loud and
control samples, respectively.}
\label{fig:control_matching}
\end{figure*}

Figure~\ref{fig:control_matching} shows the matching quality and global
properties of the final \Civ- and \Mgii-based samples.
Figure~\ref{fig:control_matching}(a) and
Figure~\ref{fig:control_matching}(b) compare the N-loud quasars with their
\Civ- and \Mgii-based controls in the
$L_{1350}$--$M_{\rm BH,C\,IV}$ and
$L_{3000}$--$M_{\rm BH,Mg\,II}$ planes, respectively. The primary \Civ-based
sample spans approximately
$45.5\lesssim\log(L_{1350}/\mathrm{erg~s^{-1}})\lesssim47.0$ and
$7.9\lesssim\log(M_{\rm BH,C\,IV}/M_\odot)\lesssim9.5$, while the smaller
\Mgii-based sample covers a narrower range. In both cases, the N-loud
quasars and their matched controls show good agreement in continuum
luminosity and virial black hole mass.

Figure~\ref{fig:control_matching}(c) and
Figure~\ref{fig:control_matching}(d) show the corresponding distributions in
the black hole mass--Eddington ratio plane for the \Civ- and \Mgii-based
samples, respectively. We calculate the Eddington ratio as
$\lambda_{\rm Edd}=L_{\rm bol}/L_{\rm Edd}$, where
$L_{\rm Edd}=1.26\times10^{38}(M_{\rm BH}/M_\odot)
\,\mathrm{erg~s^{-1}}$. For the \Civ-based sample, we adopt
$L_{\rm bol}=3.81L_{1350}$, while for the \Mgii-based sample we use
$L_{\rm bol}=5.15L_{3000}$
\citep{Richards2006,Shen2011,Rakshit2020}. The \Civ-based sample spans
approximately $-0.8\lesssim\log\lambda_{\rm Edd}\lesssim0.6$, with the
N-loud quasars and their matched controls occupying similar ranges.

\section{Emission-Line Measurements and Metallicity Estimates}
\label{sec:fitting}

\subsection{Emission-line Fitting}
\label{sec:line_fitting}

Measurements of broad UV emission lines in AGNs are often complicated by
severe line blending, making accurate line flux measurements challenging. In
practice, two approaches are commonly used to measure quasar emission-line
fluxes \citep{Nagao2006}. One is to integrate the line flux above a locally
defined continuum without explicitly modeling the line profile
\citep[e.g.,][]{Vanden2001}, while the other is to fit the emission line with
one or more appropriate functions, such as Gaussians or Lorentzians
\citep[e.g.,][]{Zheng1997}. The former is sensitive to the placement of the
local continuum and has limited ability to separate strongly blended
features. The latter provides greater flexibility in modeling complex line
profiles and separating blended components, but multi-component fits can
suffer from strong parameter degeneracies, making the decomposition
non-unique, particularly for heavily blended lines.

Here, we follow the emission-line fitting scheme of \citet{Nagao2006}, which
has subsequently been widely used in studies of BLR metallicity
\citep[e.g.,][]{Matsuoka2011,Xu2018,Lai2022}. All spectra are fitted over the
rest-frame wavelength range of 1150-2000~\AA. We first model the underlying
UV continuum after masking bad pixels and regions strongly affected by
absorption or spectral artifacts. The continuum is described by a single
power law,
\begin{equation}
F_{\lambda,\mathrm{cont}} = F_{\mathrm{cont},0}
\left(\frac{\lambda}{1450~\mathrm{\mathring{A}}}\right)^{\gamma},
\label{eq:continuum}
\end{equation}
where $F_{\mathrm{cont},0}$ and $\gamma$ represent the normalization and
power-law slope, respectively. The continuum is constrained primarily by two
relatively line-free windows at 1445-1455~\AA\ and 1973-1983~\AA. When
necessary, two additional windows at 1320-1325~\AA\ and 1370-1380~\AA\ are
used to further constrain the continuum shape. We do not include an Fe~II
pseudo-continuum component because our fitting range (1150-2000~\AA) lies
blueward of the strongest Fe~II emission complex at 2200-3000~\AA. Over the
wavelength range considered here, Fe~II emission is comparatively weak and is
expected to have only a limited effect on the strong UV lines analyzed in this
work \citep{Vestergaard2001}. The best-fitting continuum is
then subtracted before the emission-line fitting.

All emission lines are fitted simultaneously within the rest-frame wavelength
intervals 1200-1290, 1360-1430, 1450-1700, 1700-1800, and
1800-1970~\AA. Each emission line is modeled with the double power-law
profile introduced by \citet{Nagao2006},
\begin{equation}
F_{\lambda} =
\left\{
\begin{array}{ll}
F_0\left(\lambda/\lambda_0\right)^{-\alpha},
& \lambda > \lambda_0,\\[4pt]
F_0\left(\lambda/\lambda_0\right)^{+\beta},
& \lambda < \lambda_0,
\end{array}
\right.
\label{eq:double_power_law}
\end{equation}
where $F_0$ and $\lambda_0$ denote the peak flux density and peak wavelength,
respectively, while $\alpha$ and $\beta$ describe the red and blue wings of
the line profile. Detailed comparisons of different line-profile
parameterizations have shown that the resulting emission-line flux ratios are
broadly consistent among reasonable fitting functions
\citep{Nagao2006,Lai2022}.

Emission lines with different degrees of ionization are known to exhibit
systematically different velocity profiles \citep[e.g.,][]{Gaskell1982,
BaskinLaor2005}. We therefore divide the emission lines into high-ionization
lines (HILs) and low-ionization lines (LILs), with an ionization potential of
40~eV adopted as the approximate boundary between the two groups. The HILs,
including N~V, O~IV], N~IV], C~IV, and He~II, share a common pair of $\alpha$
and $\beta$, while the LILs, including Si~II, Si~IV, O~III], N~III], Al~II,
Al~III, Si~III], and C~III], share another pair. This coupling reduces the degeneracy
in decomposing strongly blended features while allowing the HILs and LILs to
retain systematically different line profiles. For Ly$\alpha$, the red-wing
index $\alpha$ is tied to that of the HILs, whereas the blue-wing index
$\beta$ is fitted independently. The flux of each emission line is obtained
by integrating its best-fitting profile. Examples of the best-fitting
spectral models are shown in Figure~\ref{fig:nloud_spectral_fit_example}.

To estimate the emission line flux uncertainties, we adopt a Monte Carlo
approach commonly used in quasar spectral analyses
\citep[e.g.,][]{Shen2019,Yang2020,Wang2021,Lai2022}. For each spectrum, we
generate 100 mock spectra by perturbing the flux at each pixel with Gaussian
random noise whose standard deviation is set by the corresponding spectral
error. The same continuum and emission line fitting procedure is applied to
each mock spectrum, with the masks defined from the original spectrum kept
fixed throughout all realizations. The final line flux of each emission line
is the median of the corresponding Monte Carlo distribution, with the lower
and upper uncertainties determined by the 16th and 84th percentiles,
respectively.

\begin{center}
\centering
\includegraphics[width=0.68\textwidth]{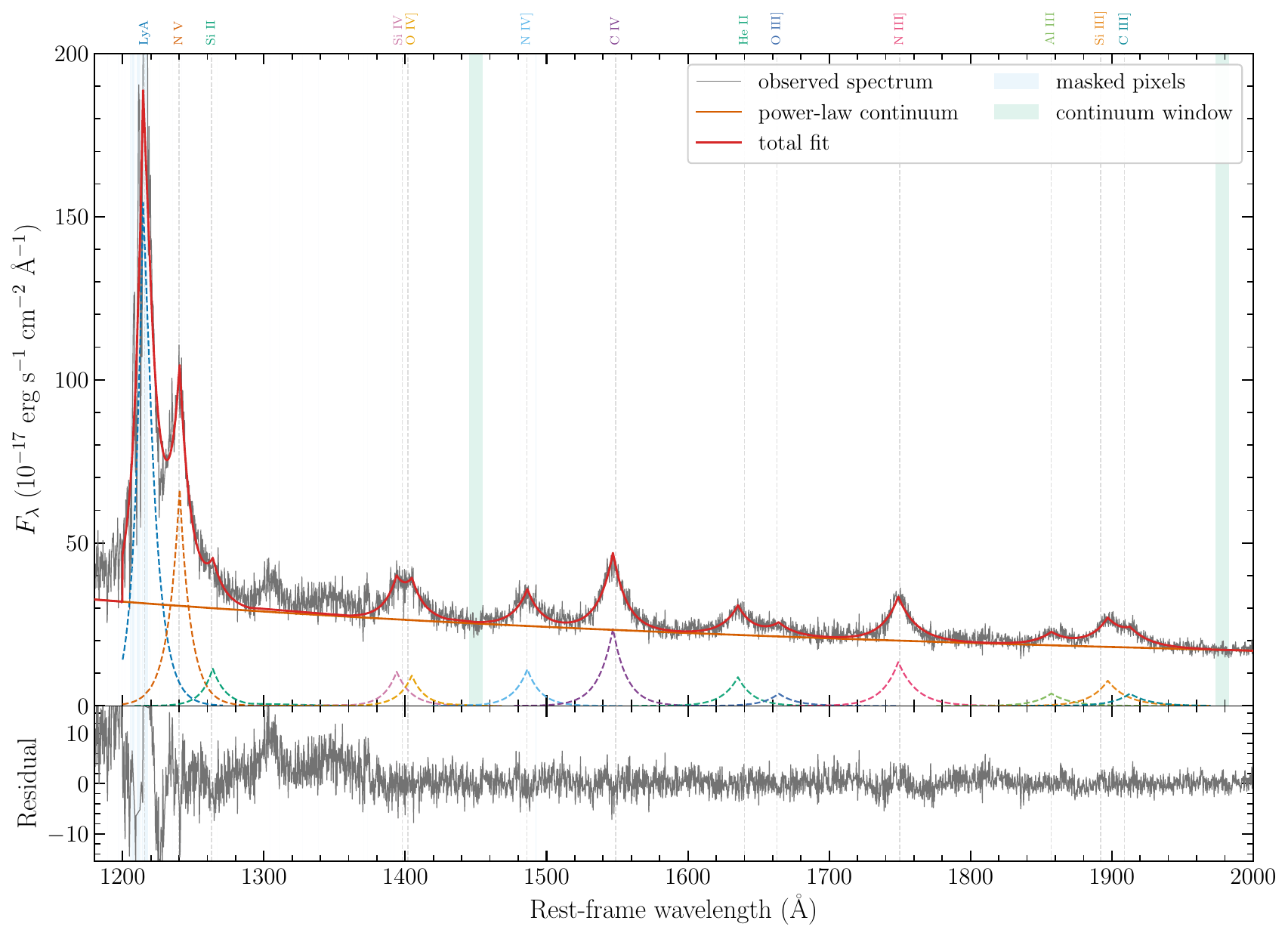}
\captionof{figure}{Representative rest-frame UV spectral decomposition for an
N-loud quasar. The gray curve shows the observed spectrum, the orange curve
denotes the fitted power-law continuum, and the red curve shows the total
best-fitting model. Colored dashed curves show the individual emission-line
components, with their identifications marked above the spectrum. Light-blue
regions indicate pixels masked due to absorption features, while the
pale-green bands mark the continuum windows used in the fit. The lower panel
shows the residuals.}
\label{fig:nloud_spectral_fit_example}
\end{center}

\subsection{Line Ratios and Metallicity}
\label{sec:line_ratios}

Broad UV emission-line ratios have long been used as diagnostics of the
chemical abundances in quasar BLRs (e.g.,
\citealt{HamannFerland1992,HamannFerland1993,Hamann2002,
Dietrich2003,Warner2003,Nagao2006,Matsuoka2011,Marziani2015,
Sameshima2017}). Theoretical calibration of these line ratios requires
photoionization modeling that accounts for the broad range of gas densities
and ionization conditions in the BLR, rather than a single-zone treatment
\citep[e.g.,][]{Davidson1977,CollinSouffrin1988}. The locally optimally
emitting cloud (LOC) framework \citep{Baldwin1995} is therefore widely used
in BLR abundance studies, in which the flux of each emission line is obtained
by integrating over an ensemble of clouds spanning a broad range of gas
density and incident ionizing flux, with each line emitted most efficiently
under its preferred physical conditions. The LOC framework has been shown to
reproduce the properties of both HILs and LILs observed in quasar spectra
\citep[e.g.,][]{KoristaGoad2000,Hamann2002,Nagao2006}.

We first consider nitrogen-sensitive abundance diagnostics, whose
metallicity dependence is based on the standard secondary-nitrogen
prescription. In this scenario, nitrogen production depends on the
pre-existing carbon and oxygen abundance, leading approximately to
$\mathrm{N/O}\propto\mathrm{O/H}$ and hence $\mathrm{N/H}\propto Z^2$ when $\mathrm{O/H}$ is used as a
tracer of the overall metallicity
\citep[e.g.,][]{Tinsley1980,HamannFerland1992,HamannFerland1993,Hamann2002}.
We adopt \Nv/\Civ\ as our primary nitrogen-sensitive diagnostic because it
has been widely used in previous BLR abundance studies and can be measured
for both the N-loud and control samples, enabling a direct comparison between
the two populations. To further investigate abundance variations within the
N-loud sample, we additionally consider \Niii/\Ciii\ and \Niii/\Oiii] as
supplementary diagnostics of N/C and N/O, respectively. Because the relevant
intercombination lines have similar ionization requirements and broadly
overlapping emitting regions, particularly for N~III] and O~III], these ratios
provide comparatively robust abundance diagnostics \citep{Hamann2002}. We do
not use
\Niv-based ratios because \Niv\ is weak or undetectable in a substantial
fraction of the N-loud quasars, preventing a homogeneous analysis across the
sample.

We also consider abundance diagnostics that do not depend on the
secondary-nitrogen prescription. We adopt
(\ion{Si}{4}+\ion{O}{4}])/\Civ\ as our primary nitrogen-independent
diagnostic, given its widespread use in previous BLR metallicity studies and
its availability for both the N-loud and control samples. Its metallicity
dependence mainly arises from changes in the thermal balance of the gas, with
the relative contribution of \Civ\ as a major coolant decreasing toward
higher metallicity. Photoionization calculations further show that this
ratio is relatively insensitive to plausible variations in the ionizing
continuum \citep[e.g.,][]{Nagao2006,Matsuoka2011}. We additionally consider
\Aliii/\Civ\ as a supplementary diagnostic. However, \Aliii\ is reliably
detected only in a subset of the N-loud quasars, limiting its use for
homogeneous statistical comparisons. Line ratios involving \Heii\ are not
included, as \Heii\ cannot be robustly measured in many individual spectra.

Following the LOC framework, we calculate the predicted emission-line fluxes
using \textup{CLOUDY} v23.01 \citep{Chatzikos2023,Gunasekera2023}, adopting
the ionizing continua and LOC model setup of \citet{Nagao2006}. We consider
metallicities of $Z/Z_\odot=0.2,\ 0.5,\ 1,\ 5,\ 10,\ 15,\ 30,\ 50,$ and
$100$. The abundances of metals are scaled linearly with $Z$, except for
nitrogen, which follows the secondary-nitrogen prescription described above.
For each metallicity and
ionizing continuum, the LOC-integrated line fluxes are used to construct the
line-ratio--metallicity relations for the diagnostics considered in this
work (see details in \citealt{Hamann2002} and \citealt{Nagao2006}). For each observed line ratio, the adopted metallicity is the mean of
the values inferred from the two ionizing continua. The observational
uncertainty is propagated through the corresponding line-ratio--metallicity
relations and averaged between the two continua, while the systematic
uncertainty is defined as one-half of the absolute difference between the
metallicity estimates obtained with the two SEDs; the two uncertainty
components are combined linearly.

\section{Results}
\label{sec:results}

In this section, we investigate the abundance properties of the N-loud
quasars using the emission-line diagnostics described in
Section~\ref{sec:fitting}. We first examine the consistency among metallicities
inferred from different diagnostics within the N-loud sample in
Section~\ref{sec:diagnostic_consistency}. We then compare the N-loud quasars
with the primary \Civ-based control sample in
Section~\ref{sec:civ_control_metallicity} and test the robustness of the
results using the independent \Mgii-based control sample in
Section~\ref{sec:mgii_control_metallicity}. Finally, we examine whether the
matched line-ratio excesses depend on black hole mass or Eddington ratio in
Section~\ref{sec:line_ratio_excess_global_properties}.

\subsection{Consistency of Metallicity Diagnostics in N-loud Quasars}
\label{sec:diagnostic_consistency}

\begin{figure*}[t]
\centering
\includegraphics[width=0.5\textwidth]{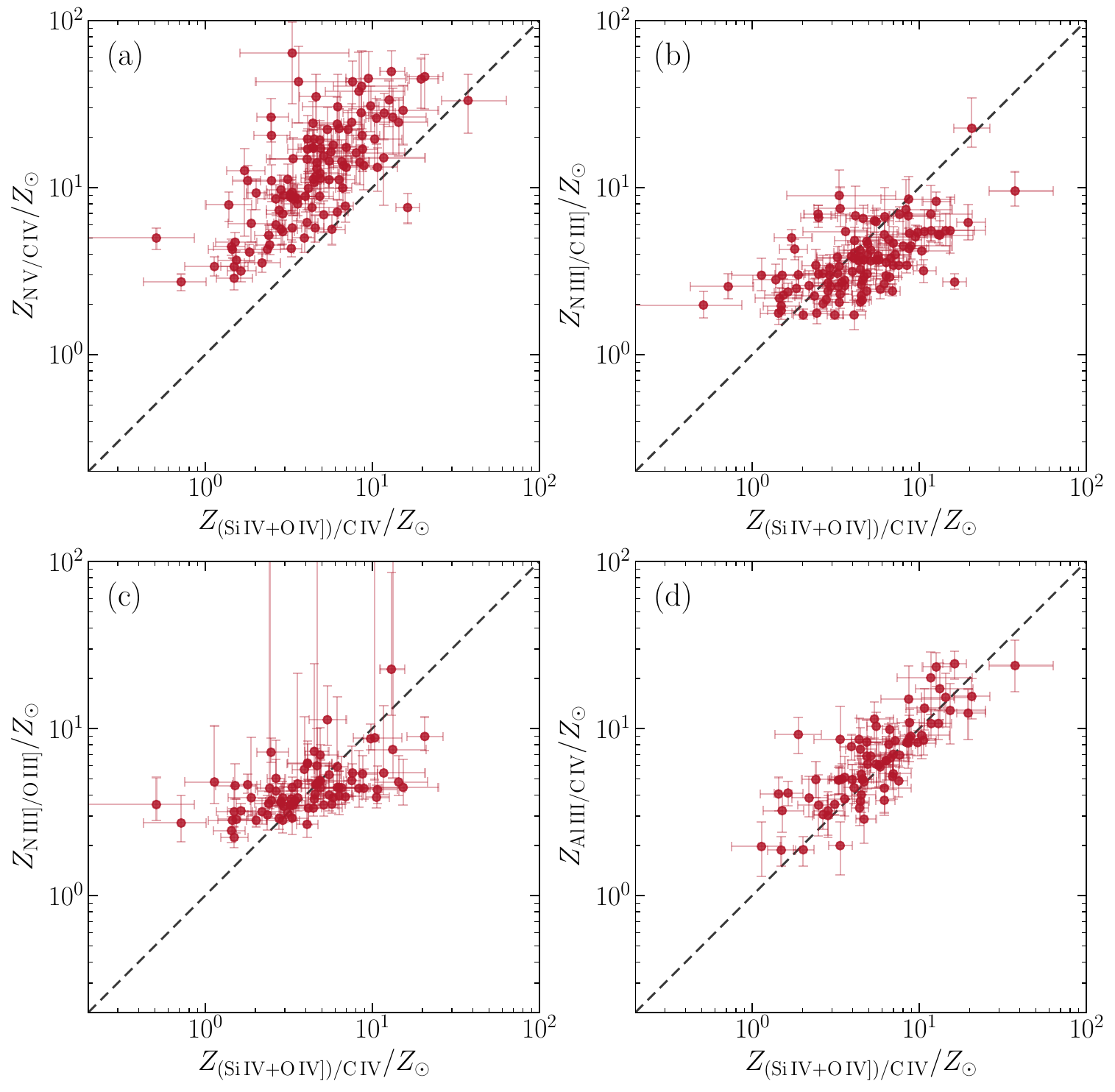}
\caption{Comparison of the metallicities inferred from different diagnostics
for the N-loud quasars. The metallicity inferred from
$(\mathrm{Si\,IV{+}O\,IV]})/\mathrm{C\,IV}$ is shown on the
horizontal axis in all panels and is compared with those inferred from
(a)~$\mathrm{N\,V}/\mathrm{C\,IV}$,
(b)~$\mathrm{N\,III]}/\mathrm{C\,III]}$,
(c)~$\mathrm{N\,III]}/\mathrm{O\,III]}$, and
(d)~$\mathrm{Al\,III}/\mathrm{C\,IV}$. The black dashed lines indicate the one-to-one
relation.}
\label{fig:diagnostics_vs_siivoiv_civ}
\end{figure*}

We first examine the consistency between metallicities inferred from
nitrogen-based and nitrogen-independent diagnostics within the N-loud sample.
Figure~\ref{fig:diagnostics_vs_siivoiv_civ}(a)--(c) compare the metallicities
inferred from $\mathrm{N\,V}/\mathrm{C\,IV}$,
$\mathrm{N\,III]}/\mathrm{C\,III]}$, and
$\mathrm{N\,III]}/\mathrm{O\,III]}$, respectively, with those inferred from
the nitrogen-independent
$(\mathrm{Si\,IV{+}O\,IV]})/\mathrm{C\,IV}$ diagnostic. In
Figure~\ref{fig:diagnostics_vs_siivoiv_civ}(a), the
$(\mathrm{Si\,IV{+}O\,IV]})/\mathrm{C\,IV}$ metallicities are
concentrated primarily between $\sim1$ and $20\,Z_\odot$, whereas the
$\mathrm{N\,V}/\mathrm{C\,IV}$ estimates span approximately
$3$-$50\,Z_\odot$. The two quantities increase together, but the
$\mathrm{N\,V}/\mathrm{C\,IV}$ metallicities are systematically higher,
typically by a factor of $\sim2.5$. A similar tendency for
$\mathrm{N\,V}/\mathrm{C\,IV}$ to yield higher metallicities has been
reported in previous quasar studies \citep{SimonHamann2010,Lai2022}. In
Figure~\ref{fig:diagnostics_vs_siivoiv_civ}(b) and (c), the
metallicities inferred from the nitrogen-based diagnostics span noticeably
narrower ranges, approximately $1$-$10\,Z_\odot$ for
$\mathrm{N\,III]}/\mathrm{C\,III]}$ and $2$-$10\,Z_\odot$ for
$\mathrm{N\,III]}/\mathrm{O\,III]}$. In particular,
$Z_{\mathrm{N\,III]}/\mathrm{O\,III]}}$ varies relatively little over the much
broader metallicity range inferred from
$(\mathrm{Si\,IV{+}O\,IV]})/\mathrm{C\,IV}$.
Figure~\ref{fig:diagnostics_vs_siivoiv_civ}(d) compares the two
nitrogen-independent diagnostics, $\mathrm{Al\,III}/\mathrm{C\,IV}$ and
$(\mathrm{Si\,IV{+}O\,IV]})/\mathrm{C\,IV}$, which yield nearly
identical metallicity ranges and show close agreement with each other. This
agreement shows that the high inferred metallicities of the N-loud quasars
are not driven solely by nitrogen-sensitive diagnostics, with nearly half of
the objects lying at $5$-$10\,Z_\odot$, extending above the typical BLR
metallicities of $\sim4$-$5\,Z_\odot$ reported in previous quasar studies
\citep{Dietrich2003,Nagao2006}.

\begin{figure*}[t]
\centering
\includegraphics[width=0.98\textwidth]{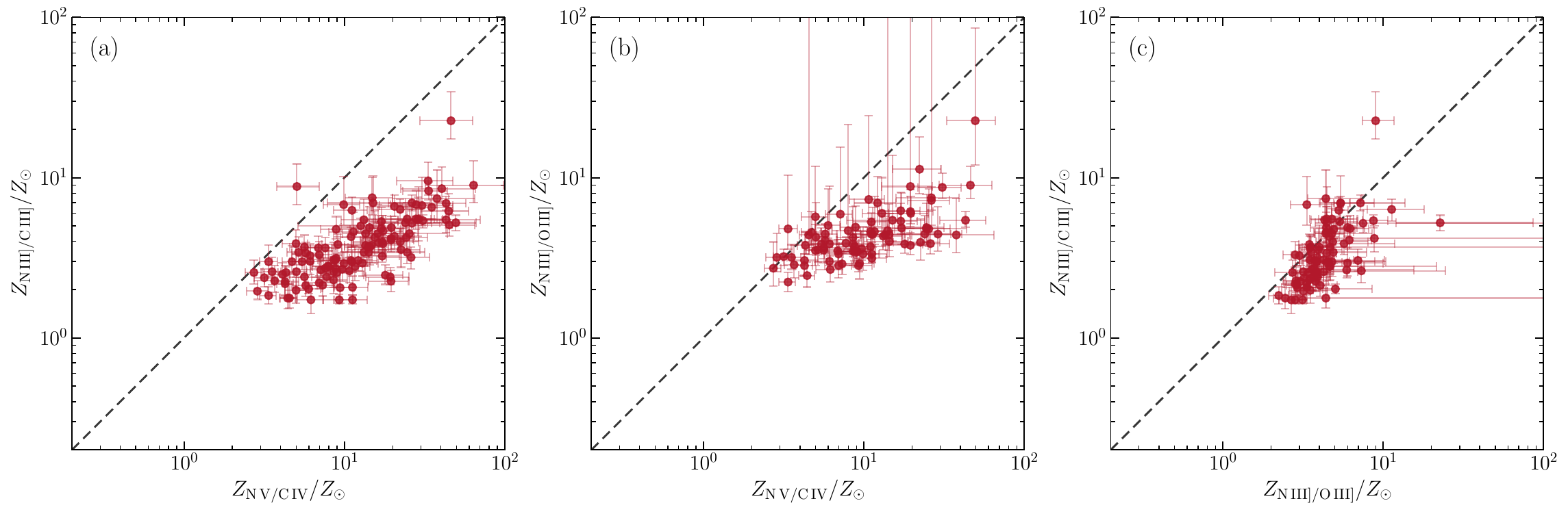}
\caption{Comparison of metallicities inferred from the nitrogen-sensitive
diagnostics for the N-loud quasars:
(a)~$\mathrm{N\,III]}/\mathrm{C\,III]}$ versus
$\mathrm{N\,V}/\mathrm{C\,IV}$,
(b)~$\mathrm{N\,III]}/\mathrm{O\,III]}$ versus
$\mathrm{N\,V}/\mathrm{C\,IV}$, and
(c)~$\mathrm{N\,III]}/\mathrm{C\,III]}$ versus
$\mathrm{N\,III]}/\mathrm{O\,III]}$. The black dashed lines indicate the one-to-one
relation.}
\label{fig:nitrogen_diagnostic_consistency}
\end{figure*}

Figure~\ref{fig:nitrogen_diagnostic_consistency} further examines the internal
consistency among the nitrogen-sensitive metallicity diagnostics. In
Figure~\ref{fig:nitrogen_diagnostic_consistency}(a) and (b), the metallicities
inferred from $\mathrm{N\,III]}/\mathrm{C\,III]}$ and
$\mathrm{N\,III]}/\mathrm{O\,III]}$ increase with
$Z_{\mathrm{N\,V}/\mathrm{C\,IV}}$, but remain systematically lower, with
$\mathrm{N\,V}/\mathrm{C\,IV}$ yielding metallicities typically about 3.5 and
2.4 times higher than $\mathrm{N\,III]}/\mathrm{C\,III]}$ and
$\mathrm{N\,III]}/\mathrm{O\,III]}$, respectively.
Figure~\ref{fig:nitrogen_diagnostic_consistency}(c), in contrast, shows much
closer agreement between the two N~III]-based diagnostics, with
$\mathrm{N\,III]}/\mathrm{O\,III]}$ yielding metallicities only about 30\%
higher than $\mathrm{N\,III]}/\mathrm{C\,III]}$. The largest discrepancy
among the nitrogen-sensitive diagnostics is therefore associated with
$\mathrm{N\,V}/\mathrm{C\,IV}$, whereas the two N~III]-based diagnostics
remain relatively consistent and still imply metallicities of
$\sim2$-$10\,Z_\odot$.

The different behavior of these three nitrogen-sensitive diagnostics likely
reflects their different sensitivities to the physical conditions and
structure of the BLR. N~V and C~IV are both high-ionization permitted lines
with very high critical densities ($\sim10^{15}\ {\rm cm^{-3}}$). N~V
preferentially traces more highly ionized gas than C~IV, so N~V/C~IV
preferentially samples highly ionized BLR gas and remains effective over a
broad range of gas densities. \citet{Hamann2002} also showed that N~V/C~IV is
sensitive to the ionization state and the adopted ionizing SED.
\citet{Temple2021} further demonstrated that substantial variations in this
ratio can occur even at fixed metallicity and that the ratio is linked to the
kinematic structure of the high-ionization BLR. These sensitivities may partly
account for the broader and systematically higher metallicity range inferred
from N~V/C~IV. By contrast, N~III], O~III], and C~III] are intercombination
lines with critical densities of only $\sim10^{10}\ {\rm cm^{-3}}$ and are
collisionally suppressed at higher densities, so they preferentially trace
lower-density BLR gas. Their ionization requirements and emitting regions are
also broadly similar, particularly for N~III] and O~III]. The more similar
physical conditions sampled by the two N~III]-based diagnostics are
consistent with their closer mutual agreement.

The different line-emitting conditions imply that the N~V- and N~III]-based
diagnostics place different weights on the BLR density--ionization structure,
which can contribute to their different inferred metallicities. Therefore,
the differences among these diagnostics alone make it difficult to establish
additional nitrogen enrichment within an individual quasar. Such an excess
is more robustly tested by comparing the same diagnostics between N-loud and
matched normal quasars, as we do in
Section~\ref{sec:civ_control_metallicity}.

\subsection{Comparison with the C IV-based Control Sample}
\label{sec:civ_control_metallicity}

To reduce the impact of the diagnostic-dependent systematics discussed in
Section~\ref{sec:diagnostic_consistency}, we compare the N-loud quasars with
controls matched in redshift, continuum luminosity, and black hole mass.
Because the same diagnostics are applied to both samples, systematic
residual differences are less likely to arise solely from line-specific
sensitivities to BLR physical conditions.

\begin{figure*}[!htbp]
\centering
\includegraphics[width=0.5\textwidth]{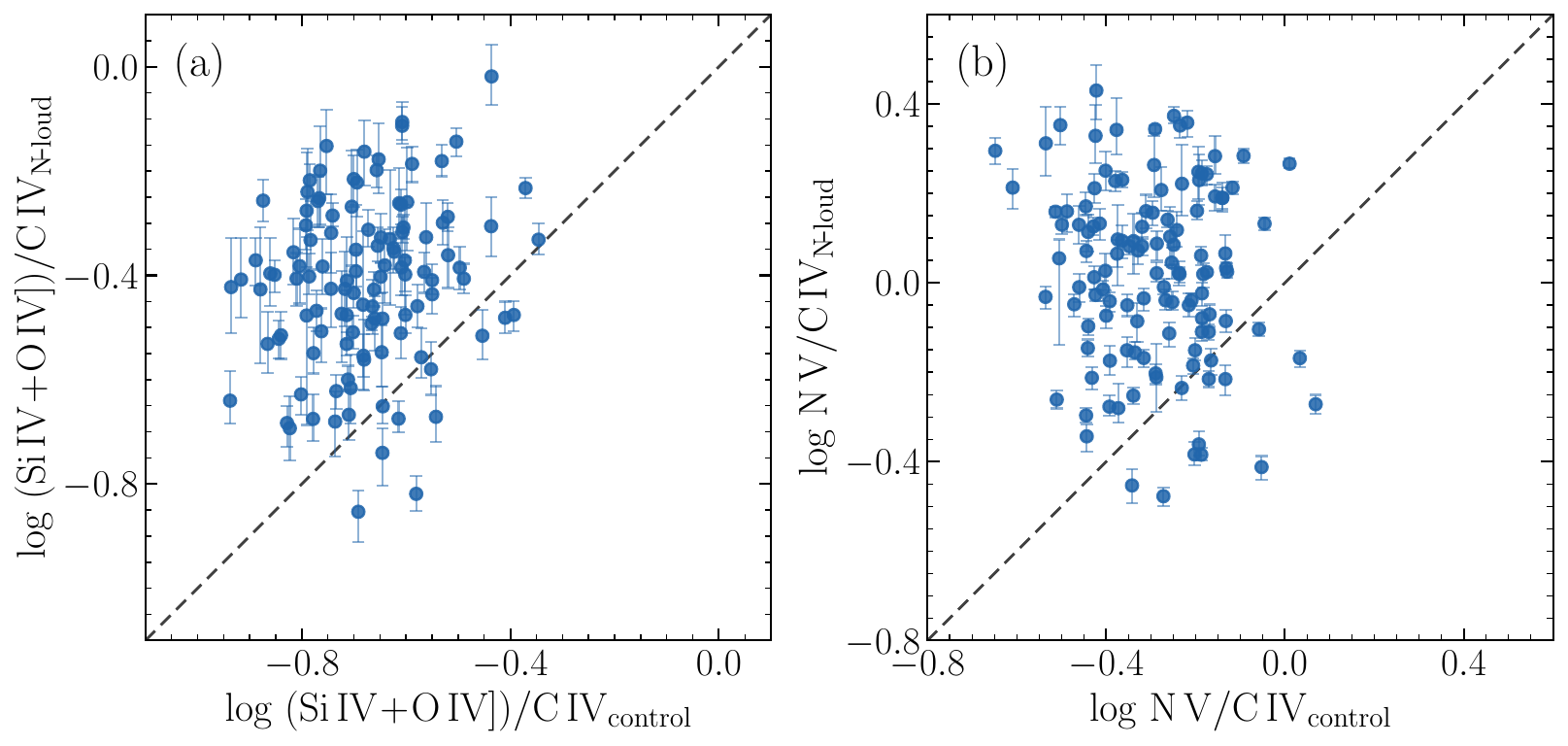}
\caption{Comparison of the broad-line ratios for the N-loud quasars
and their \Civ-based matched controls: (a)~$(\mathrm{Si\,IV{+}O\,IV]})/
\mathrm{C\,IV}$ and (b)~$\mathrm{N\,V}/\mathrm{C\,IV}$. For each N-loud
quasar, the control value is taken as the median line ratio of its
corresponding matched-control set. The dashed lines indicate the one-to-one
relation.}
\label{fig:civ_control_matched_line_ratios}
\end{figure*}

We first examine the directly
observed line ratios, as metallicity estimates from photoionization modeling
depend on several assumptions. Figure~\ref{fig:civ_control_matched_line_ratios}
compares the $(\mathrm{Si\,IV{+}O\,IV]})/\mathrm{C\,IV}$ and
$\mathrm{N\,V}/\mathrm{C\,IV}$ ratios of each N-loud quasar with the median
ratios of its corresponding \Civ-based controls. For
$(\mathrm{Si\,IV{+}O\,IV]})/\mathrm{C\,IV}$, the control quasars are
concentrated primarily at $\log$ line ratios of approximately $-0.9$ to
$-0.4$, whereas the N-loud quasars lie mainly between $-0.7$ and $-0.1$.
For $\mathrm{N\,V}/\mathrm{C\,IV}$, the controls are concentrated at
approximately $-0.5$ to $-0.1$, while the N-loud quasars span roughly $-0.3$
to $0.4$. In both comparisons, about 90\% of the N-loud quasars lie above the
one-to-one relation, indicating a clear systematic enhancement relative to
their matched controls.

\begin{figure*}[!htbp]
\centering
\includegraphics[width=0.5\textwidth]{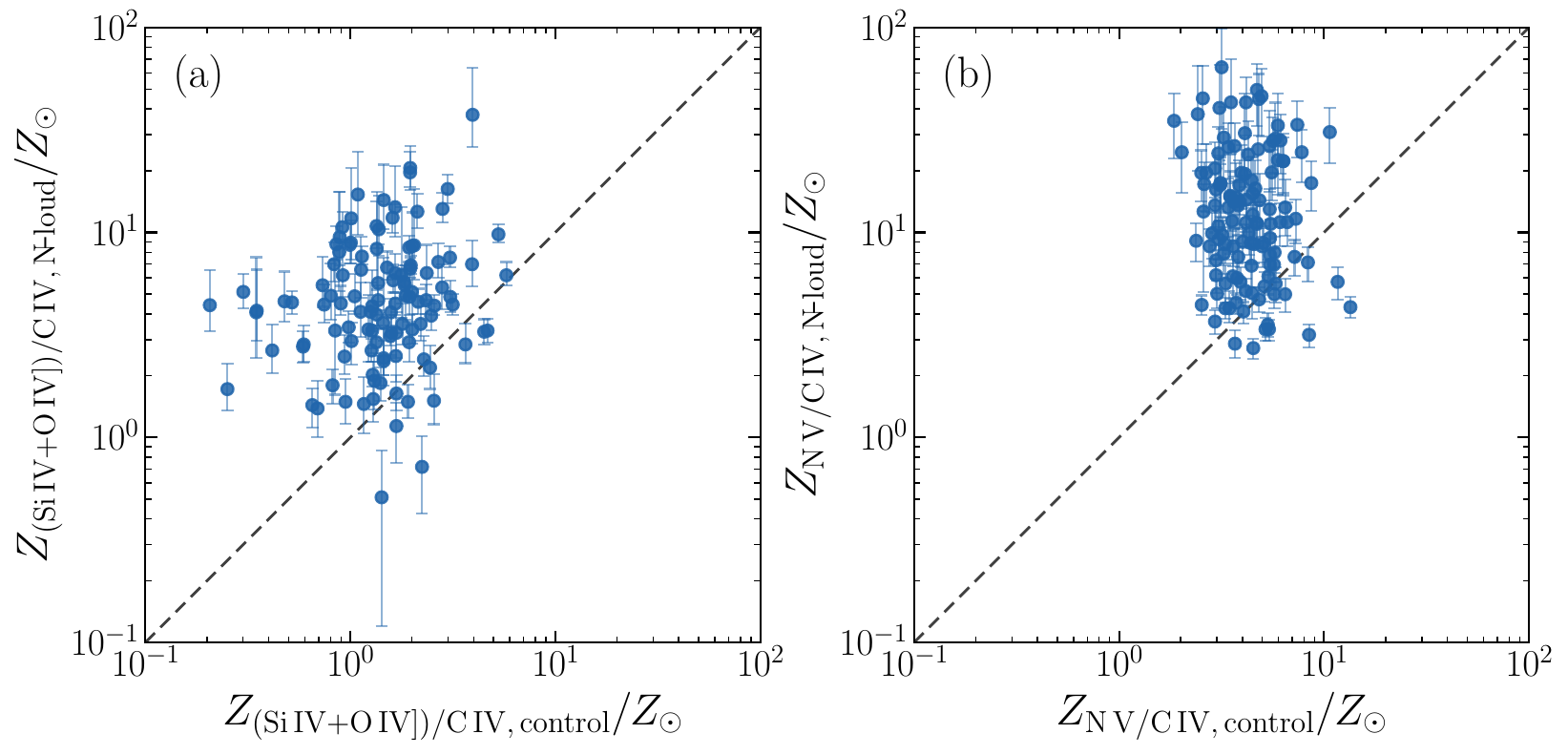}
\caption{Comparison of the metallicities inferred for the N-loud
quasars and their \Civ-based matched controls using
(a)~$(\mathrm{Si\,IV{+}O\,IV]})/\mathrm{C\,IV}$ and
(b)~$\mathrm{N\,V}/\mathrm{C\,IV}$. For each N-loud quasar, the control
metallicity is taken as the median metallicity of its corresponding
matched-control set. The dashed lines indicate the one-to-one relation.}
\label{fig:civ_control_matched_metallicities}
\end{figure*}

Figure~\ref{fig:civ_control_matched_metallicities}
shows that this offset persists after conversion to metallicity. For
$(\mathrm{Si\,IV{+}O\,IV]})/\mathrm{C\,IV}$, the control metallicities are
concentrated primarily between $\sim0.5$ and $5\,Z_\odot$, whereas the N-loud
quasars lie mainly between $\sim1$ and $12\,Z_\odot$. For
$\mathrm{N\,V}/\mathrm{C\,IV}$, the controls are concentrated at
$\sim2$-$10\,Z_\odot$, while the N-loud quasars extend over
$\sim3$-$50\,Z_\odot$. Most N-loud quasars again lie above the one-to-one
relation in both diagnostics. Thus, the N-loud quasars show systematically
higher inferred metallicities than their matched controls in both the
nitrogen-independent and nitrogen-based diagnostics. In particular, the
excess shown in the nitrogen-independent diagnostic provides strong evidence
that the N-loud population is overall more metal rich, rather than being
characterized by nitrogen enhancement alone.

\begin{figure*}[!htbp]
\centering
\includegraphics[width=\textwidth]{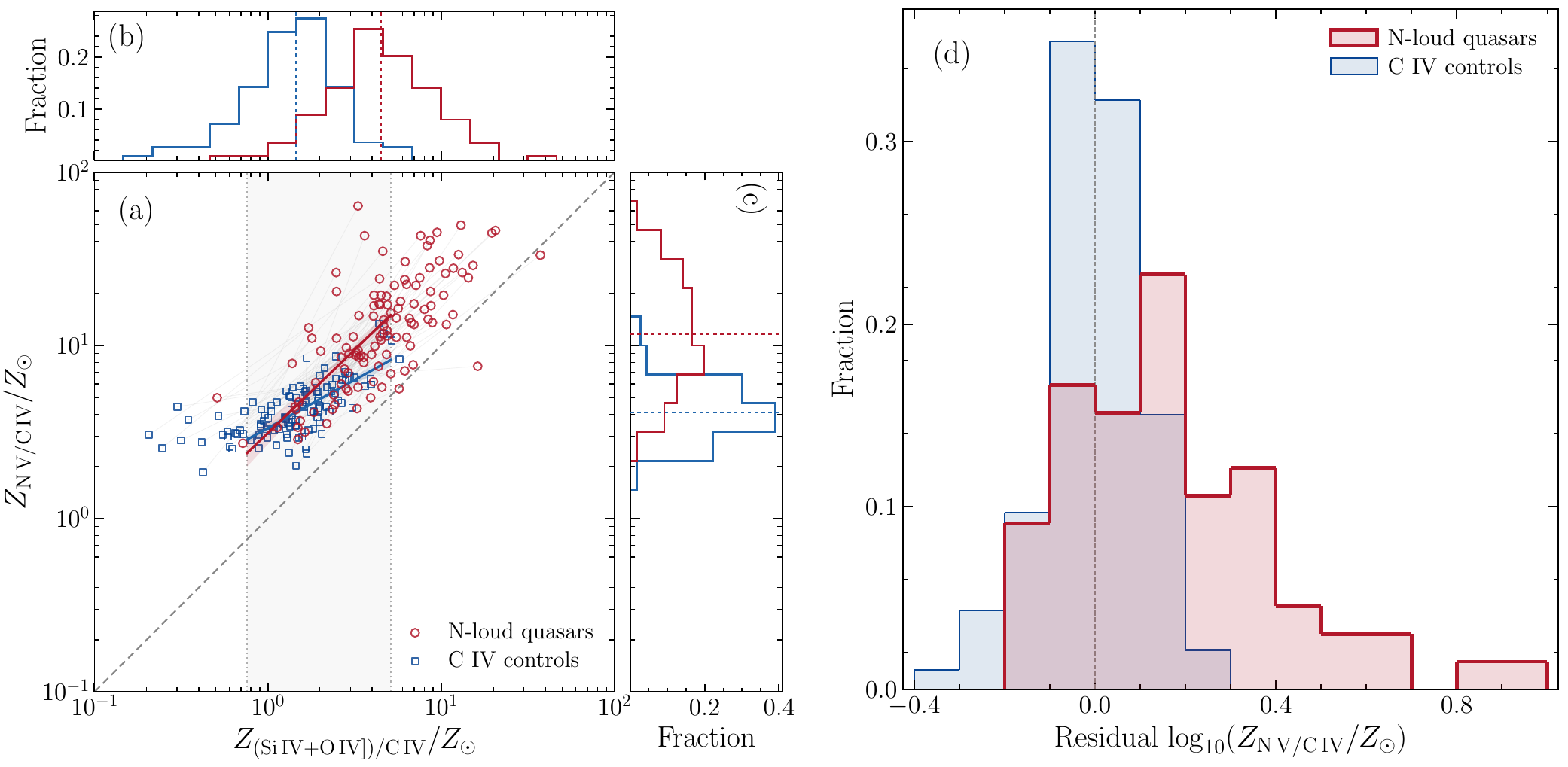}
\caption{Comparison of the metallicities inferred from
$(\mathrm{Si\,IV{+}O\,IV]})/\mathrm{C\,IV}$ and
$\mathrm{N\,V}/\mathrm{C\,IV}$ for the N-loud quasars and their \Civ-based
matched controls. In panel (a), red open circles show the N-loud quasars and
blue open squares show the median metallicities of their corresponding
matched-control sets; faint gray segments connect each N-loud quasar to its
matched-control median. The gray dashed line indicates the one-to-one
relation. Panels (b) and (c) show the marginal distributions,
with dashed lines marking the corresponding medians. The vertical dotted
lines in panel (a) delimit the metallicity range used for the regression
analysis, and the red and blue solid lines show the best-fitting relations
for the N-loud and control samples, respectively; shaded regions indicate
the 68\% confidence intervals. Panel (d) shows the residual distributions
of the N-loud quasars and matched-control medians, using the best-fitting
relation for the C IV-based control sample as a common baseline for both
populations. For each point, the residual is defined as
$r=y-(a_{\rm C}+b_{\rm C}x)$, where $a_{\rm C}$ and $b_{\rm C}$ are the
intercept and slope of the C IV-based control relation, respectively.
Positive residuals indicate higher N V/C IV-based metallicities than
predicted by the C IV-based control relation at the same
(Si IV+O IV])/C IV-based metallicity. Here,
$x=\log_{10}[Z_{(\mathrm{Si\,IV{+}O\,IV]})/\mathrm{C\,IV}}/Z_\odot]$
and $y=\log_{10}[Z_{\mathrm{N\,V}/\mathrm{C\,IV}}/Z_\odot]$.
Only points within the common fitting range are included in panel (d). The gray dashed
line marks zero residual. In panels (b)--(d), each histogram bin shows
the fraction of its respective sample.}
\label{fig:civ_control_diagnostic_comparison}
\end{figure*}

Figure~\ref{fig:civ_control_diagnostic_comparison} further compares the joint
distributions of the two metallicity diagnostics for the N-loud quasars and
the control sample. The median
$Z_{(\mathrm{Si\,IV{+}O\,IV]})/\mathrm{C\,IV}}$ and
$Z_{\mathrm{N\,V}/\mathrm{C\,IV}}$ values for the N-loud quasars are
approximately $5\,Z_\odot$ and $12\,Z_\odot$, respectively, compared with
approximately $1.5\,Z_\odot$ and $4\,Z_\odot$ for the controls. To minimize
boundary effects, we restrict the regression analysis to the overlapping
range between the 1st and 99th percentiles of the two samples,
$0.76<Z_{(\mathrm{Si\,IV{+}O\,IV]})/\mathrm{C\,IV}}/Z_\odot<5.13$. Both
samples show a positive relation between the two metallicity estimates, but
the relation is substantially steeper for the N-loud quasars, with slopes of
0.96 and 0.56 for the N-loud and control samples, respectively. We fit both
samples simultaneously using an ordinary least-squares linear model that
allows separate intercepts and slopes for the two groups. A two-sided
$t$-test of the slope difference yields $p=0.011$. The two
relations are similar at the lower-metallicity end but diverge progressively
toward higher metallicity, indicating that the relative N V excess becomes
more pronounced as the metallicity inferred from the nitrogen-independent
diagnostic increases. This differential trend suggests additional nitrogen
enrichment beyond the overall metallicity enhancement in the N-loud quasars,
a result discussed further in Section~\ref{sec:overall_enrichment}.

To quantify the overall offset relative to the control relation, we compare
the residual distributions of the two samples within the same common
metallicity range, as shown in Figure~\ref{fig:civ_control_diagnostic_comparison}(d).
For each point, the residual is the difference between its logarithmic
N V/C IV-based metallicity and that predicted by the best-fitting control
relation at the same reference metallicity. Both samples are evaluated
relative to this same control relation. The median residual of the N-loud
sample exceeds that of the controls by $\Delta r_{\rm med}=0.142$ dex.
To estimate the uncertainty, we treat each N-loud quasar and its
matched-control set as a single unit in the bootstrap resampling. In each
realization, we refit the control relation and recalculate the difference in
median residuals. The resulting 95\% confidence interval is $[0.060, 0.220]$
dex. This result supports an overall positive offset in the N V/C IV-based
metallicity of the N-loud sample relative to the control relation within
this range, complementing the slope comparison above.

\subsection{Robustness to the Mg II-based Control Sample}
\label{sec:mgii_control_metallicity}

\begin{figure*}[!htbp]
\centering
\includegraphics[width=0.5\textwidth]{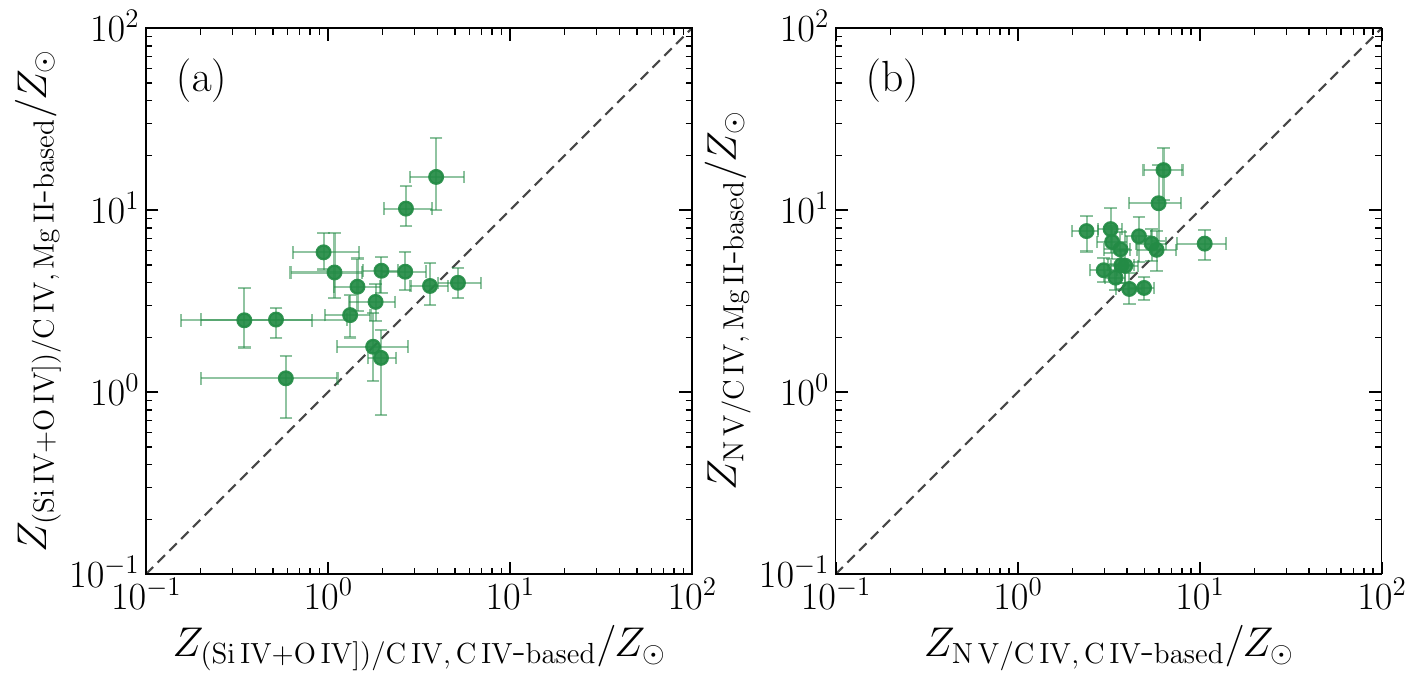}
\caption{Comparison between the metallicities of the \Civ-based
and \Mgii-based matched-control baselines for N-loud quasars with both
control sets available. Panels (a) and (b) show the metallicities inferred
from $(\mathrm{Si\,IV{+}O\,IV]})/\mathrm{C\,IV}$ and
$\mathrm{N\,V}/\mathrm{C\,IV}$, respectively. Each point compares the
median metallicities of the corresponding \Civ-based and \Mgii-based
control sets. The dashed lines indicate the one-to-one relation.}
\label{fig:civ_mgii_control_baseline}
\end{figure*}

\begin{figure*}[!htbp]
\centering
\includegraphics[width=0.5\textwidth]{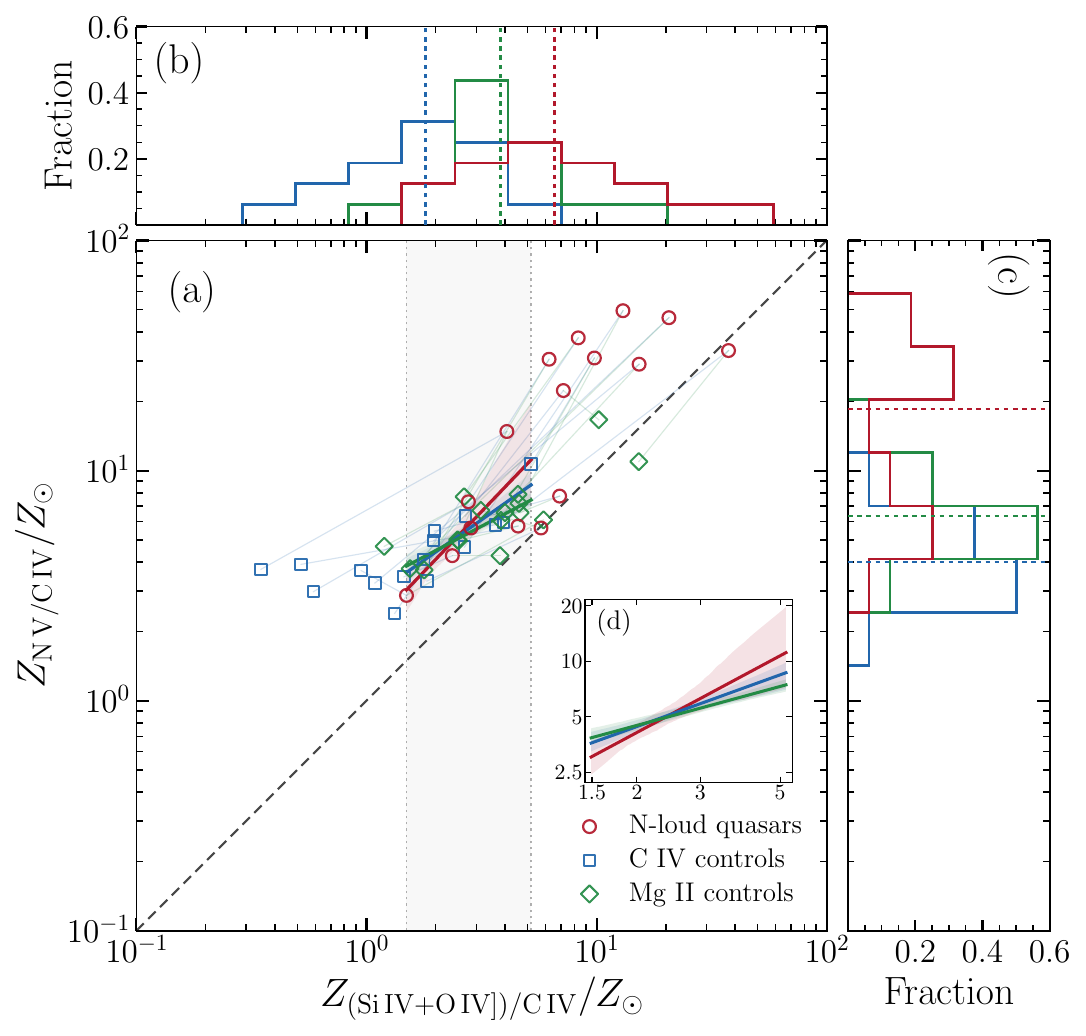}
\caption{\label{fig:nloud_civ_mgii_control_diagnostic_comparison}%
Metallicity comparison for the N-loud quasars and the
two independently constructed matched-control samples. Panel (a) shows the
metallicities inferred from
$(\mathrm{Si\,IV{+}O\,IV]})/\mathrm{C\,IV}$ and
$\mathrm{N\,V}/\mathrm{C\,IV}$ for the N-loud quasars (red circles), the
median values of their \Civ-based control sets (blue squares), and those of
their \Mgii-based control sets (green diamonds). Faint line segments connect
each N-loud quasar to its corresponding control medians, and the gray dashed
line marks the one-to-one relation. Panels (b) and (c) show the normalized
marginal distributions, with dashed lines marking the sample medians. The
colored solid lines and shaded regions show the best-fitting relations and
their 68\% confidence intervals over the metallicity range used for the
regressions; panel (d) provides an enlarged view of these fits.}
\end{figure*}

Given that \Civ-based black hole mass estimates can be affected by non-virial
motions and outflow-related line structure
\citep{ShenLiu2012,Coatman2016,Coatman2017}, we further use an independently
constructed \Mgii-based control sample as a robustness test. We select the 16
N-loud quasars for which both \Civ- and \Mgii-based control sets are
available. Figure~\ref{fig:civ_mgii_control_baseline} shows that the \Mgii-based
controls generally yield somewhat higher metallicity baselines than the
corresponding \Civ-based controls, particularly for the nitrogen-independent
$(\mathrm{Si\,IV{+}O\,IV]})/\mathrm{C\,IV}$ diagnostic. Nevertheless,
Figure~\ref{fig:nloud_civ_mgii_control_diagnostic_comparison} shows that the
N-loud quasars remain systematically more metal rich than both control samples
in both diagnostics. The median
$Z_{(\mathrm{Si\,IV{+}O\,IV]})/\mathrm{C\,IV}}$ values are approximately
$7\,Z_\odot$, $2\,Z_\odot$, and $4\,Z_\odot$ for the N-loud sample and the
\Civ- and \Mgii-based control samples, respectively, while the corresponding
$Z_{\mathrm{N\,V}/\mathrm{C\,IV}}$ medians are approximately $19\,Z_\odot$,
$4\,Z_\odot$, and $6\,Z_\odot$. Over the metallicity range jointly covered
by the three samples, the fitted slopes are 1.04, 0.70, and 0.53 for the
N-loud sample and the \Civ- and \Mgii-based control samples, respectively. Although the
small number of available objects limits the statistical significance of
these slopes, the steeper relation for the N-loud quasars is consistent with
the trend seen in Figure~\ref{fig:civ_control_diagnostic_comparison}. Thus,
the main conclusions of systematically higher metallicity and additional
nitrogen enrichment in the N-loud quasars remain unchanged when the control
sample is constructed using \Mgii-based black hole masses.

\subsection{Dependence of Line-Ratio Excesses on Black Hole Mass and Eddington Ratio}
\label{sec:line_ratio_excess_global_properties}

\begin{figure*}[!htbp]
\centering
\includegraphics[width=0.5\textwidth]{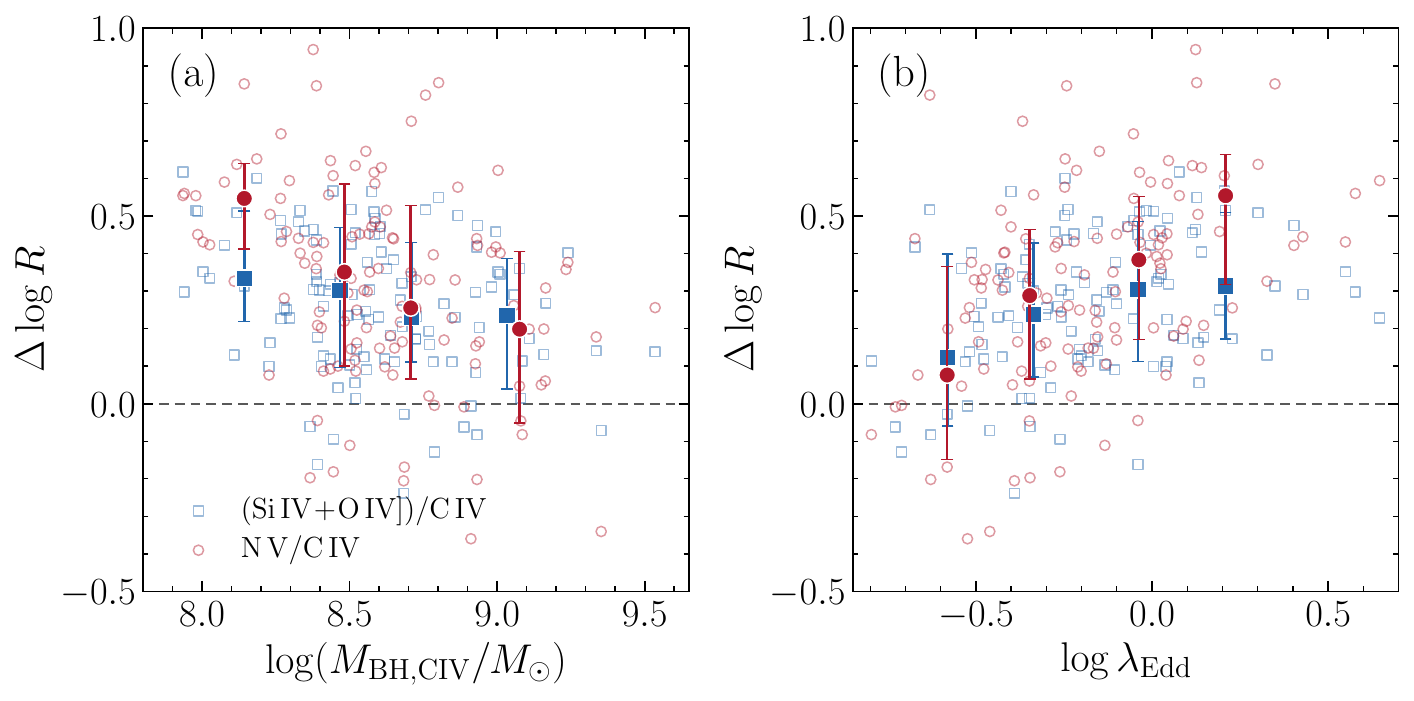}
\caption{Matched line-ratio excesses of the N-loud quasars as a
function of black hole mass and Eddington ratio. Panels (a) and (b) show
$\Delta\log R$ versus the \Civ-based black hole mass and Eddington ratio,
respectively. Blue squares represent
$(\mathrm{Si\,IV{+}O\,IV]})/\mathrm{C\,IV}$, while red circles represent
$\mathrm{N\,V}/\mathrm{C\,IV}$. Open symbols show individual N-loud
quasars; the larger filled symbols and vertical error bars indicate the
binned medians and corresponding 16th-84th percentile ranges. The
horizontal dashed line marks $\Delta\log R=0$, corresponding to no excess
relative to the matched controls.}
\label{fig:line_ratio_excess_global_properties}
\end{figure*}

We further examine whether the line-ratio excesses of the N-loud quasars
relative to their matched controls depend on black hole mass or Eddington
ratio. For each diagnostic, we define the matched excess as the difference
between the logarithmic line ratio of an N-loud quasar and the median value of
its \Civ-based matched controls,
$\Delta\log R_i=\log R_{{\rm N\mbox{-}loud},i}-
{\rm median}_j(\log R_{{\rm control},ij})$.
Figure~\ref{fig:line_ratio_excess_global_properties}(a) shows the matched
excess as a function of black hole mass. For both
$(\mathrm{Si\,IV{+}O\,IV]})/\mathrm{C\,IV}$ and
$\mathrm{N\,V}/\mathrm{C\,IV}$, the median excess remains positive across
the full black hole mass range. The
$(\mathrm{Si\,IV{+}O\,IV]})/\mathrm{C\,IV}$ excess changes only mildly,
decreasing from $\sim0.3$ dex at the low-mass end to $\sim0.2$ dex at the
high-mass end, whereas the $\mathrm{N\,V}/\mathrm{C\,IV}$ excess decreases
from $\sim0.6$ to $\sim0.2$ dex. Thus, the enhancement in both line ratios
persists across the sampled black hole mass range, with the excess appearing
stronger toward lower black hole masses, particularly for
$\mathrm{N\,V}/\mathrm{C\,IV}$.

Figure~\ref{fig:line_ratio_excess_global_properties}(b) shows the matched
excess as a function of Eddington ratio. The median excess remains positive
for both diagnostics across the full Eddington-ratio range. The
$(\mathrm{Si\,IV{+}O\,IV]})/\mathrm{C\,IV}$ excess increases from $\sim0.1$
to $\sim0.3$ dex, while the $\mathrm{N\,V}/\mathrm{C\,IV}$ excess rises more
strongly, from $\sim0.1$ to $\sim0.6$ dex, suggesting that the additional
$\mathrm{N\,V}/\mathrm{C\,IV}$ enhancement may become more pronounced at
higher Eddington ratios. A qualitatively similar Eddington-ratio trend for
\Nv-based line ratios was reported by \citet{Matsuoka2011}.
In addition, the black hole masses of highly accreting
quasars may be systematically overestimated by conventional single-epoch
virial estimators that do not account for accretion-state effects
\citep{DuWang2019,Maithil2022,Pan2025}. If a similar bias affects our
\Civ-based estimates, the affected objects would shift toward lower black
hole masses in Figure~\ref{fig:line_ratio_excess_global_properties}(a) and
higher Eddington ratios in
Figure~\ref{fig:line_ratio_excess_global_properties}(b), potentially altering
the apparent trends in both panels. Given
these potential systematic effects, the
substantial object-to-object scatter, and the overlap among the percentile
ranges, we do not assign statistical significance to these apparent trends.
Nevertheless, the positive binned medians across the full ranges of black
hole mass and Eddington ratio show that the line-ratio excesses are not
confined to a particular parameter regime.

\section{Discussion}
\label{sec:discussion}

\subsection{Overall Metallicity and Relative Nitrogen Enrichment}
\label{sec:overall_enrichment}

Our matched-sample analysis provides direct evidence that the BLRs of N-loud
quasars are overall more metal rich than those of otherwise similar normal
quasars. The nitrogen-independent $(\mathrm{Si\,IV{+}O\,IV]})/\mathrm{C\,IV}$
diagnostic yields a median metallicity approximately three times higher than
that of the matched \Civ-based controls. The agreement with \Aliii/\Civ\ and
the qualitatively similar result obtained with the \Mgii-based control sample
provide additional support for this interpretation. The abundance pattern of
N-loud quasars therefore cannot be explained by selective nitrogen enhancement
alone.

The overall metallicity enhancement, however, does not fully account for the
behavior of the nitrogen-sensitive diagnostic. The adopted calibrations already
include the increase in relative nitrogen abundance expected from secondary
nitrogen production \citep{HamannFerland1993,Hamann2002,Nagao2006}. If N-loud
quasars simply represented the high-metallicity extension of the normal-quasar
abundance sequence, they would be expected to follow approximately the same
relation between the nitrogen-based and nitrogen-independent metallicity
estimates. Instead, within the common reference-metallicity range, the N-loud
quasars show both a significantly steeper relation and an overall positive
residual offset relative to the controls. This population-level departure,
rather than the discrepancy between the two diagnostics within individual
objects, provides evidence for additional relative nitrogen enrichment beyond
that associated with the overall metallicity enhancement. The steeper relation
further suggests that the nitrogen excess becomes more pronounced toward the
metal-rich end of the N-loud population.

Previous studies have generally emphasized either high overall metallicity
\citep{Baldwin2003,Batra2014} or enhanced nitrogen abundance relative to other
elements \citep{Jiang2008} as the origin of the unusual nitrogen emission in
N-loud quasars. By combining a nitrogen-independent metallicity reference with
matched normal-quasar controls analyzed using the same diagnostics and
photoionization prescriptions, our analysis distinguishes these effects more
directly. The results show that the two interpretations are not mutually
exclusive, with the N-loud quasars being overall more metal rich while also
exhibiting an additional nitrogen excess relative to normal quasars at
comparable reference metallicity.

This abundance pattern suggests that relative nitrogen abundance is at least
partially decoupled from overall metallicity in N-loud quasars. Increasing a
single metallicity parameter while retaining the normal-quasar abundance
pattern does not fully describe their inferred chemical properties. Models of
nuclear enrichment must therefore explain both the elevated overall metallicity
and the additional nitrogen enrichment relative to normal quasars at the same
reference metallicity. The possible origins of this abundance pattern and its
connection to black hole fueling are discussed in Section 5.2.

\subsection{Enrichment Scenarios and Their Connection to Black Hole Accretion}

N-loud quasars have been reported to differ systematically from normal quasars
in their accretion properties. \citet{Batra2014} found narrower \Civ\ emission
lines and lower \Civ-based virial black hole masses at comparable continuum
luminosities. Using both H$\beta$- and \Civ-based estimates,
\citet{Matsuoka2017} likewise found lower black hole masses and higher Eddington
ratios in a sample of 12 N-loud quasars. \citetalias{Zhai2026PaperI} extended this result to a
substantially larger DESI sample, showing that N-loud quasars have systematically
narrower \Civ\ and \Mgii\ lines, lower virial black hole masses, and higher
Eddington ratios than normal quasars matched in redshift and UV luminosity.
These results support an empirical association between the N-loud phenomenon
and accretion state, raising the question of whether the unusual abundance
pattern itself is related to black hole accretion.

Figure~\ref{fig:line_ratio_excess_global_properties} provides an initial test
of this connection by examining whether the line-ratio excesses of N-loud
quasars relative to their matched controls vary with accretion state. Because
the controls are matched in continuum luminosity and virial black hole mass,
the N-loud quasars and their controls occupy similar Eddington-ratio ranges.
The excesses in both $(\mathrm{Si\,IV{+}O\,IV]})/\mathrm{C\,IV}$ and
$\mathrm{N\,V}/\mathrm{C\,IV}$ remain positive across the full sampled range
of $\lambda_{\rm Edd}$, while the $\mathrm{N\,V}/\mathrm{C\,IV}$ excess shows
a stronger apparent increase toward higher Eddington ratio. Given the
substantial scatter and overlap among the percentile ranges, we do not assign
statistical significance to this trend. A larger sample will be required to
determine whether the $\mathrm{N\,V}/\mathrm{C\,IV}$ excess is indeed more
closely linked to accretion state than the
$(\mathrm{Si\,IV{+}O\,IV]})/\mathrm{C\,IV}$ excess. Such a differential
dependence would provide an additional constraint on enrichment scenarios for
N-loud quasars.

A frequently invoked interpretation for N-loud quasars links their nitrogen
enrichment to delayed mass return from intermediate-mass asymptotic giant
branch (AGB) stars. Intermediate-mass AGB stars are important contributors to
nitrogen production through processes such as hot-bottom burning and return
their processed material through relatively slow stellar winds
\citep{KarakasLugaro2016}. Stellar mass loss from evolved nuclear populations
has also been proposed as a source of gas for SMBH accretion and as a possible
explanation for the delay between nuclear star formation and AGN activity
\citep{Davies2007}. This picture has been invoked in the context of N-loud
quasars to connect enhanced nitrogen emission with black hole fueling
\citep{Matsuoka2011,Matsuoka2017}. In this scenario, nitrogen enrichment may
precede a short-lived episode of rapid black hole accretion. In light of the
tentative trend shown in Figure~\ref{fig:line_ratio_excess_global_properties}(b),
such delayed stellar mass return could contribute to the additional nitrogen
enrichment, provided that the returned material mixes with gas that is already
substantially metal rich.

In addition to delayed stellar mass return, nitrogen enrichment may also arise
from processes operating within the AGN accretion environment itself. Recent
theoretical studies have explored the evolution of stars embedded in AGN
accretion disks, where the surrounding gas can substantially modify their
growth and subsequent evolution. Sustained accretion, internal mixing, and mass
loss can alter the stellar evolutionary pathways, while stellar winds and
ejecta from later evolutionary stages return chemically processed material
directly to the disk \citep{Cantiello2021,Jermyn2022,Huang2023,Xu2026}. Such
models provide a framework in which both the overall metallicity and relative
abundance pattern of the accreting gas can be modified by stellar processing.
In the ``immortal-star'' regime, continuous accretion and efficient mixing can
replenish hydrogen in the stellar core and sustain CNO burning, processing C
and O into N through sustained CNO cycling and returning N-enhanced material to
the disk through stellar winds \citep{Jermyn2022,Xu2026}. Tidal disruption
events provide an additional transient channel, in which CNO-processed stellar
material can be injected into the nuclear environment and temporarily produce
nitrogen-rich spectra \citep{Kochanek2016,Liu2018}.

These scenarios operate on different spatial and temporal scales, and the
present data do not distinguish among them. Any successful enrichment model
for N-loud quasars must reproduce both the elevated overall BLR metallicity and
the additional nitrogen enrichment, while remaining consistent with their
observed accretion properties. More flexible abundance modeling, in which
nitrogen is allowed to vary independently of overall metallicity, will be
important for quantifying the relative nitrogen enhancement and testing
specific enrichment scenarios.

\subsection{Connections to Nitrogen-Enriched Systems at High Redshift}

JWST observations have identified several high-redshift systems with nitrogen
abundances elevated relative to their overall metallicities. GN-z11 is a
prominent example, showing a strongly enhanced N/O ratio at a substantially
subsolar oxygen abundance \citep{Cameron2023}, and similar abundance anomalies
have been reported in other galaxies at $z\gtrsim5$
\citep{Isobe2023,MarquesChaves2024,Topping2024}. These systems occupy a very
different metallicity regime from the N-loud quasars studied here, which are
already strongly metal enriched. Nevertheless, both populations show that
relative nitrogen abundance is not uniquely determined by overall metallicity.
In the N-loud quasars, the nitrogen excess is superimposed on an already
metal-rich BLR, whereas many of the high-redshift systems exhibit enhanced N/O
at substantially lower overall metallicity.

Several high-redshift nitrogen-enhanced systems further suggest that the
anomalous nitrogen emission can be associated with dense nuclear gas. In GS
3073 at $z=5.55$, the nitrogen abundance inferred from the UV-emitting gas is
substantially higher than that inferred from optical lines, indicating strong
stratification between different gas components \citep{Ji2024}. Stacked spectra
of broad-line AGNs at $z=4$-7 likewise show enhanced N/C and N/O together with
high gas densities, whereas the corresponding narrow-line AGN stack does not
show comparable nitrogen features \citep{Isobe2025}. These measurements do not
establish that the nitrogen-rich gas originates in the BLR, but they indicate
that strong nitrogen enhancement can be confined to dense gas in the nuclear
environment. This behavior is qualitatively consistent with localized nuclear
enrichment, although abundance measurements from different gas phases cannot
be compared directly.

Nitrogen enhancement is also emerging as a recurring spectroscopic property in
at least a subset of little red dots (LRDs). CANUCS-LRD-z8.6 at $z=8.63$, an LRD
with spectroscopic evidence for an AGN, shows strong N IV] emission indicative
of nitrogen enrichment despite its low overall metallicity
\citep{Morishita2026}. Two LRDs at $z=6.68$ and 8.35 studied by
\citet{Papovich2026} likewise show nitrogen enhancement together with strongly
stratified, high-density gas, while recent ultra-deep spectroscopy from the
SPURS program finds nitrogen enhancement in all four LRDs in the sample
\citep{Tang2026}. The latter systems also exhibit unusually dense nuclear gas
and, in some cases, broad high-ionization emission. Although these results do
not yet establish nitrogen enhancement as a universal property of LRDs, they
strengthen the empirical association between nitrogen-rich gas and compact,
dense nuclear environments at high redshift.

The physical connection between these high-redshift systems and N-loud quasars
remains uncertain. Their different metallicity regimes suggest that the
dominant enrichment pathways need not be the same. At low metallicity and early
cosmic times, rapidly rotating massive stars, Wolf--Rayet stars, very massive
or supermassive stars, and stellar interactions in dense clusters have been
proposed as sources of nitrogen-rich material
\citep{Cameron2023,Isobe2025,Nandal2025}. By contrast, the high overall
metallicities of N-loud quasars permit a larger contribution from delayed
stellar enrichment and chemical processing in mature nuclear environments, as
discussed in Section 5.2. A possible common feature is therefore the ability of
dense nuclear environments to produce, retain, or concentrate chemically
processed material, rather than a single enrichment mechanism shared by all of
these systems. The N-loud quasars studied here may thus provide a
high-metallicity counterpart to the nitrogen-enhanced systems identified at
much earlier cosmic epochs, rather than representing a direct evolutionary
extension of the same population.

\section{Conclusions}\label{sec:sum}

To investigate the BLR abundance properties associated with the strong
nitrogen emission in N-loud quasars, we analyze a high-quality sample of 121
N-loud quasars selected from the DESI N-loud quasar catalog presented in Paper
I, spanning $2.13 \leq z \leq 3.90$ with complete rest-frame UV spectral
coverage over 1150-2000~$\mathrm{\mathring{A}}$. We uniformly measure the
principal broad UV emission lines and infer BLR metallicities from
nitrogen-sensitive and nitrogen-independent line ratios using
\textup{CLOUDY}-based photoionization models. For comparison with otherwise
similar normal quasars, we construct control samples matched in redshift,
continuum luminosity, and virial black hole mass, using \Civ-based masses for
the primary analysis and \Mgii-based masses as an independent robustness test.
Our main results are summarized as follows.

\textbf{(i)} Within the N-loud quasar sample, metallicities inferred from
different broad UV diagnostics show substantial systematic differences. Among
the nitrogen-sensitive diagnostics, \Nv/\Civ\ yields the broadest and highest
metallicities, spanning $Z\sim3$-$50\,Z_\odot$, whereas the two \Niii-based
diagnostics are largely confined to $\sim1$-$10\,Z_\odot$ and show much closer
agreement with each other. Compared with the nitrogen-independent
(\ion{Si}{4}+\ion{O}{4}])/\Civ\ diagnostic, \Nv/\Civ\ also yields
systematically higher metallicities, with most
(\ion{Si}{4}+\ion{O}{4}])/\Civ-based estimates lying at
$Z\sim1$-$20\,Z_\odot$. By contrast, the two nitrogen-independent diagnostics,
(\ion{Si}{4}+\ion{O}{4}])/\Civ\ and \Aliii/\Civ, show close agreement.

\textbf{(ii)} Relative to matched normal quasars, N-loud quasars show
systematically higher overall BLR metallicities together with evidence for
additional nitrogen enrichment. In both
(\ion{Si}{4}+\ion{O}{4}])/\Civ\ and \Nv/\Civ, about 90\% of the N-loud
quasars exceed the medians of their \Civ-based controls. The median
metallicities are $\sim5\,Z_\odot$ and $\sim1.5\,Z_\odot$ for the N-loud quasars
and \Civ-based controls, respectively, based on
(\ion{Si}{4}+\ion{O}{4}])/\Civ, and $\sim12\,Z_\odot$ and $\sim4\,Z_\odot$
based on \Nv/\Civ. More importantly, within the overlapping range in
(\ion{Si}{4}+\ion{O}{4}])/\Civ-based metallicity, the relation between
\Nv/\Civ-based and (\ion{Si}{4}+\ion{O}{4}])/\Civ-based metallicities is
steeper for the N-loud quasars than for the \Civ-based controls, with slopes of
0.96 and 0.56, respectively; the difference is statistically significant
($p=0.011$), supporting additional nitrogen enrichment beyond the overall metal
enrichment. The smaller \Mgii-based control sample shows the same qualitative
abundance trends.

\textbf{(iii)} The line-ratio excesses of the N-loud quasars relative to their
\Civ-based matched controls remain positive across the sampled ranges of black
hole mass and Eddington ratio, indicating that the abundance enhancements are
not restricted to a particular region of parameter space. Moreover, the
\Nv/\Civ\ excess shows a tentative trend toward larger values at lower black
hole masses and higher Eddington ratios.

Overall, these results indicate that relative nitrogen abundance and overall
metallicity may be at least partially decoupled in the BLRs of N-loud quasars
and that this abundance pattern is not fully described by the standard
secondary-nitrogen scaling alone. Larger N-loud quasar samples with reliable
\Mgii-based black hole masses and complete UV abundance diagnostics will enable
more stringent tests of whether the relative nitrogen excess systematically
increases with overall metallicity and of its dependence on black hole
accretion properties. Further modeling of possible enrichment channels will
help constrain the physical origin of this abundance pattern.
\begin{acknowledgments}

This study is supported by the National Natural Science Foundation of China (NSFC) under grant No. 12588202, the Strategic Priority Research Program of the Chinese Academy of Sciences under grant No. XDB1160103, the National Key R\&D Program of China under grant Nos. 2024YFA1611903, 2023YFE0107800, and 2024YFA1611601, and the CAS Project for Young Scientists in Basic Research under grant No. YSBR-092.

This research used data obtained with the Dark Energy Spectroscopic Instrument (DESI). DESI construction and operations is managed by the Lawrence Berkeley National Laboratory. This material is based upon work supported by the \href{https://www.energy.gov/}{U.S. Department of Energy}, Office of Science, Office of High-Energy Physics, under Contract No. DE--AC02--05CH11231, and by the National Energy Research Scientific Computing Center, a DOE Office of Science User Facility under the same contract. Additional support for DESI was provided by the \href{https://www.nsf.gov/}{U.S. National Science Foundation} (NSF), Division of Astronomical Sciences under Contract No. AST-0950945 to the NSF's National Optical-Infrared Astronomy Research Laboratory; the \href{https://stfc.ukri.org/}{Science and Technology Facilities Council of the United Kingdom}; the \href{https://www.moore.org/}{Gordon and Betty Moore Foundation}; the \href{https://www.hsfoundation.org/}{Heising-Simons Foundation}; the \href{https://www.cea.fr/}{French Alternative Energies and Atomic Energy Commission} (CEA); the \href{https://secihti.mx/}{National Council of Humanities, Science and Technology of Mexico} (CONAHCYT); the \href{http://www.mineco.gob.es/}{Ministry of Science and Innovation of Spain} (MICINN), and by the DESI Member Institutions: \href{https://www.desi.lbl.gov/collaborating-institutions}{www.desi.lbl.gov/collaborating-institutions}. The DESI collaboration is honored to be permitted to conduct scientific research on I'oligam Du'ag (Kitt Peak), a mountain with particular significance to the \href{https://www.tonation-nsn.gov/}{Tohono O'odham Nation}. Any opinions, findings, and conclusions or recommendations expressed in this material are those of the author(s) and do not necessarily reflect the views of the U.S. National Science Foundation, the U.S. Department of Energy, or any of the listed funding agencies.

\end{acknowledgments}

\appendix
\renewcommand{\theHequation}{appendix.\thesection.\arabic{equation}}

\section{Virial Black Hole Mass Measurements}
\label{app:civ_control}

\subsection{C IV-based Black Hole Masses}
\label{app:civ_fitting}
\label{app:civ_bh_mass}

The \Civ-based virial black hole masses are used to construct the primary
control sample. We apply the same mass-measurement procedure to the
N-loud quasars and quasars in the normal-quasar parent sample. The
single-epoch black hole masses are estimated using the \Civ-based calibration of
\citet{Vestergaard2006},
\begin{equation}
\log\left(\frac{M_{\rm BH}}{M_\odot}\right)=
6.66+0.53\log\left(
\frac{L_{1350}}{10^{44}~\mathrm{erg~s^{-1}}}
\right)
+2\log\left(
\frac{\mathrm{FWHM}_{\mathrm{C\,IV}}}
{1000~\mathrm{km~s^{-1}}}
\right),
\label{eq:civ_vp06}
\end{equation}
where $L_{1350}\equiv\lambda L_\lambda(1350~\mathrm{\mathring{A}})$ is the
monochromatic continuum luminosity at rest-frame
1350~$\mathrm{\mathring{A}}$, and $\mathrm{FWHM}_{\mathrm{C\,IV}}$ is the
full width at half maximum of the broad \Civ\ emission line.

The continuum is fitted with a power law using the line-free windows at
1445-1455 and 1973-1983~$\mathrm{\mathring{A}}$, and $L_{1350}$ is measured
from the best-fitting continuum at 1350~$\mathrm{\mathring{A}}$. After
subtracting the continuum, the \Civ\ emission over
1500-1600~$\mathrm{\mathring{A}}$ is fitted with three broad Gaussian
components, following standard quasar spectral-fitting procedures
\citep{Vestergaard2006,Shen2011,Rakshit2020}. Pixels affected by
obvious absorption features are masked during the fit. We measure
$\mathrm{FWHM}_{\mathrm{C\,IV}}$ directly from the total best-fitting \Civ\
profile rather than from any individual Gaussian component. No additional
correction for the \Civ\ blueshift is applied.

Uncertainties in the black hole mass estimates are evaluated using Monte
Carlo simulations. We generate 100 mock spectra for each N-loud quasar and 50
for each quasar in the normal-quasar parent sample by perturbing the observed
flux at each pixel according to a Gaussian distribution with a standard
deviation given by the corresponding spectral uncertainty. Each mock spectrum
is then refitted following the same procedure as for the observed spectrum,
yielding new measurements of $L_{1350}$,
$\mathrm{FWHM}_{\mathrm{C\,IV}}$, and $M_{\rm BH}$. The statistical
uncertainty in $M_{\rm BH}$ is estimated from the central 68\% interval of the
resulting mass distribution. These uncertainties account for the propagation
of spectral measurement errors but do not include the intrinsic systematic
scatter of the single-epoch virial mass calibration.

\subsection{Mg II-based Black Hole Masses}
\label{app:mgii_bh_mass}

The \Mgii-based virial black hole masses are used to construct an independent
control sample for the subset of quasars with rest-frame spectral
coverage over 1150-3000~$\mathrm{\mathring{A}}$. We apply the same
mass-measurement procedure to the N-loud quasars and quasars in the
normal-quasar parent sample. The single-epoch black hole masses are estimated
using the \Mgii-based calibration of \citet{Vestergaard2009},
\begin{equation}
\log\left(\frac{M_{\rm BH}}{M_\odot}\right)=
6.86+0.50\log\left(
\frac{L_{3000}}{10^{44}~\mathrm{erg~s^{-1}}}
\right)
+2\log\left(
\frac{\mathrm{FWHM}_{\mathrm{Mg\,II}}}
{1000~\mathrm{km~s^{-1}}}
\right),
\label{eq:mgii_vo09}
\end{equation}
where $L_{3000}\equiv\lambda L_\lambda(3000~\mathrm{\mathring{A}})$ is the
monochromatic continuum luminosity at rest-frame
3000~$\mathrm{\mathring{A}}$, and $\mathrm{FWHM}_{\mathrm{Mg\,II}}$ is the
full width at half maximum of the broad \Mgii\ emission line.

The continuum is modeled with a power law together with UV \Feii\ emission
represented by the I~Zw~1 template of \citet{Vestergaard2001}. The fit is
constrained using the rest-frame windows at 1345-1350, 1445-1455,
1700-1705, 1973-1983, 2150-2400, 2480-2675, and
2900-3000~$\mathrm{\mathring{A}}$, while the \Mgii-dominated
2700-2900~$\mathrm{\mathring{A}}$ region is excluded. No separate
Balmer-continuum component is included. $L_{3000}$ is measured from the
best-fitting power-law continuum at 3000~$\mathrm{\mathring{A}}$.

After subtracting the power-law plus \Feii\ pseudo-continuum, the \Mgii\
emission over 2700-2900~$\mathrm{\mathring{A}}$ is fitted with two broad
Gaussian components. These components provide a flexible representation of
the blended \Mgii\ profile and are not interpreted as the two physical
members of the doublet. We measure $\mathrm{FWHM}_{\mathrm{Mg\,II}}$
directly from the total best-fitting \Mgii\ profile rather than from any
individual Gaussian component.

Uncertainties in the \Mgii-based black hole masses are estimated following
the same Monte Carlo procedure as for the \Civ-based measurements.

\begingroup
\subsection{Comparison of C IV- and Mg II-based Black Hole Masses}
\label{app:civ_mgii_bh_mass_comparison}

To assess the consistency of the two black hole mass estimators, we compare
the \Civ- and \Mgii-based black hole masses for 16 N-loud quasars and
109 control quasars with both measurements available. The control quasars
comprise 52 objects from the \Civ-based matched-control sets and 57 from
the \Mgii-based matched-control sets associated with these 16 N-loud quasars.
Figure~\ref{fig:civ_mgii_bh_mass_comparison} presents this comparison, with
the dashed line indicating the one-to-one relation. The logarithmic mass
difference, defined as
$\log M_{\rm BH,C\,IV}-\log M_{\rm BH,Mg\,II}$, has a median of $0.049$ dex
and a standard deviation of $0.179$ dex for the N-loud quasars. The
corresponding values for the controls are $0.038$ dex and $0.319$ dex,
respectively. Although individual objects exhibit appreciable scatter,
particularly among the controls, the small median offsets in both samples
indicate that the two estimators yield broadly consistent black hole masses,
with no appreciable systematic offset between them.

\begin{center}
\includegraphics[width=0.5\textwidth]{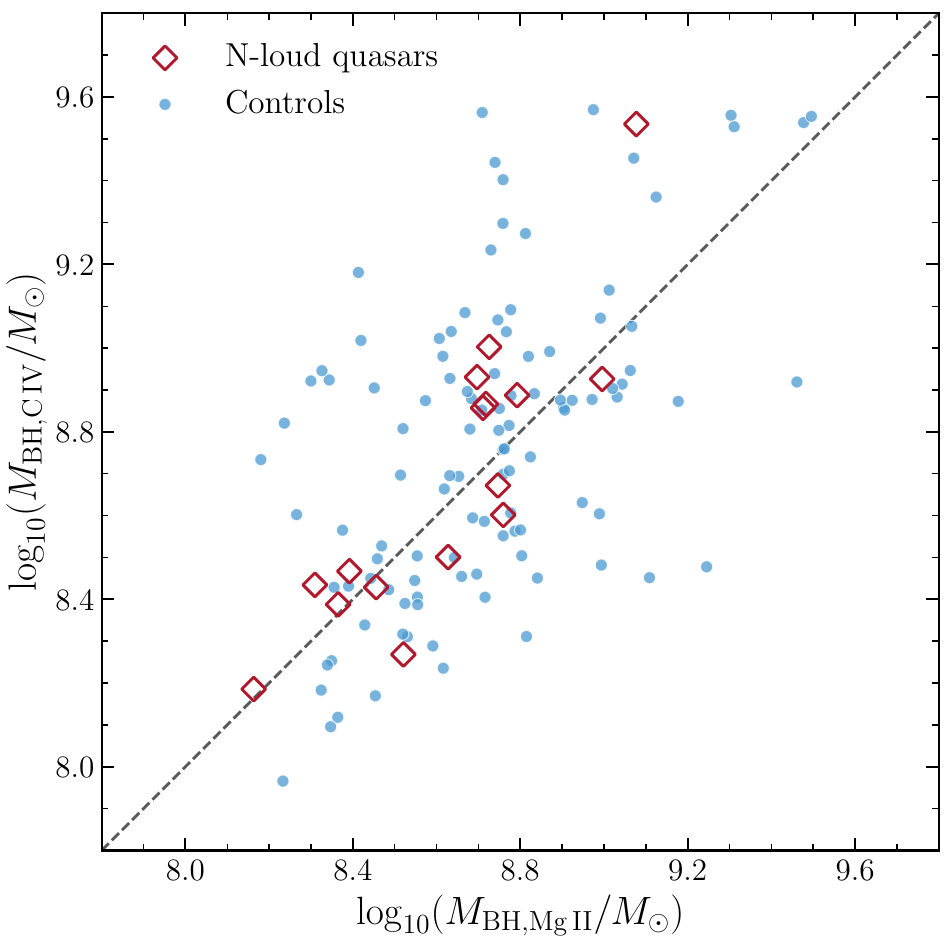}
\captionof{figure}{Comparison of C IV- and Mg II-based black hole
masses for 16 N-loud quasars (red open diamonds) and their 109 matched control
quasars (blue filled circles) with both measurements available. The dashed
line indicates the one-to-one relation.}
\label{fig:civ_mgii_bh_mass_comparison}
\end{center}
\endgroup

\bibliography{sample701}{}
\bibliographystyle{aasjournalv7}

\end{document}